\documentclass{article}
\usepackage[utf8]{inputenc}
\usepackage[a4paper,left=3cm,right=3cm,top=3cm,bottom=3cm]{geometry}
\usepackage{amsmath}
\usepackage{amssymb}
\usepackage{graphicx}
\usepackage{caption}
\usepackage{subcaption}
\usepackage{esint}
\usepackage{enumitem}
\usepackage{multirow}
\usepackage{graphicx}
\usepackage{hhline}
\usepackage{hyperref}
\usepackage{centernot}
\usepackage{comment}
\usepackage{cite}
\hypersetup{
  colorlinks   = true,    
  urlcolor     = blue,    
  linkcolor    = blue,    
  citecolor    = black     
}
\newcommand{\KaTie}{{\sc Ka\hspace{-0.2ex}Tie}}

\title{Forward $\gamma$+jet production in proton-proton and proton-lead collisions at LHC within the FoCal calorimeter acceptance}

\author{
Ishita Ganguli$^a$,
Andreas van Hameren$^b$,
Piotr Kotko$^a$,
Krzysztof Kutak$^b$
\\ \\
$^a${\it AGH University Of Krakow,}\\ 
{\it Faculty of Physics and Applied Computer Science,} \\ 
{\it al. Mickiewicza 30, 30-059 Krak\'ow, Poland} \\ \\
$^b${\it Institute of Nuclear Physics, Polish Academy of Sciences,}\\ 
{\it Radzikowskiego 152, 31-342 Krak\'ow, Poland } \\
}

\date{}

\begin{document}

\maketitle

\begin{abstract}
  Using the small-$x$ Improved Transverse Momentum Dependent factorization (ITMD), which, for a general two-to-two massless scattering can be proved within the Color Glass Condensate (CGC) theory for transverse momenta of particles greater than the saturation scale, we provide predictions for  isolated forward photon and jet production in proton-proton and proton-nucleus collisions within the planned ALICE FoCal detector acceptance. We study azimuthal correlations, $p_T$ spectra, as well as normalized ratios of proton-proton cross sections for different energies.
  The only TMD distribution needed for that process is the "dipole" TMD gluon distribution, which in our computations is based on HERA data and undergoes momentum space BK evolution equation with DGLAP corrections and Sudakov resummation. 
  We conclude, that the process provides an excellent probe of the dipole TMD gluon distribution in saturation regime.
\end{abstract}

\section{Introduction}
\label{sec:intro}

Despite Quantum Chromodynamics (QCD) seems to be the correct theory of strong interactions, there are still puzzles that await direct experimental verification. One of those is the phenomenon of the ``gluon saturation'' \cite{Gribov:1984tu}, i.e.\ the regime of parton densities where the perturbative splitting of gluons is balanced by the gluon recombination, so that the gluon density stops its power-like growth. Actually the significance of the saturation domain goes well beyond the perturbative ``partonic'' language and is often treated via the effective high energy description called the Color Glass Condensate (CGC) (for a review see eg.~\cite{Gelis:2010nm,Kovchegov:2012mbw}). 
Regardless of whether it is CGC formalism, or perturbative Pomeron splitting, the saturation is a prediction of QCD and is expected to occur at asymptotically large energies. 
Although there are numerous experimental hints that it has been already observed -- see the direct experimental data \cite{STAR:2006dgg,PHENIX:2011puq,STAR:2021fgw} as well as various phenomenological saturation-based approaches confronted with HERA, RHIC and LHC data \cite{Golec-Biernat:1998zce,Lappi:2012nh,Albacete:2018ruq,Stasto:2018rci,vanHameren:2019ysa,Benic:2022ixp}, there always exists an alternative explanation, either within the collinear factorization (supplemented by other effects like MPI or resummations), or by other possible nuclear effects eg.~\cite{Frankfurt:2011cs}.

Potential discovery of saturation and its detailed properties are part of the Electron Ion Collider (EIC) program \cite{Accardi:2012qut}. However, before the EIC starts operating, there is still an enormous potential in the LHC program. So far there were no LHC experiments directly aiming at studies of saturation physics, which is most likely to be seen only at the very forward rapidity region in collisions of protons and nuclei. This gap will be filled by the FoCal calorimeter of ALICE collaboration \cite{ALICECollaboration:2719928,ALICE:2023fov}. The new detector will be able to measure with very good resolution jets and photons in the very forward region, with pseudorapidity ultimately covering the interval $3.4 < \eta < 5.8$. 

In that context, the process we would like to study, and which has been investigated earlier in context of saturation in \cite{Rezaeian:2012wa,Jalilian-Marian:2012wwi} is the production of an isolated on-shell photon and a jet in hadron-hadron collision
\begin{equation}
    p(P_B)+A(P_A) \rightarrow \gamma(k_1) + J(k_2)+X \,,
    \label{eq:main_process}
\end{equation}
where $A$ may be either a proton or a nucleus target, which in our study will always be lead.
$P_B$ and $P_{A}$ are the incoming hadron and nucleus momenta and $k_1$, $k_2$ are the momenta of the photon and the reconstructed jet, respectively.
We are interested in the final state kinematics that probes the target $A$ at small longitudinal momentum fractions $x_A$. More precisely, $x_A=k_A\cdot P_B/P_A\cdot P_B$, where $k_A$ is the four momentum exchanged between the partons extracted from the target $A$ and the rest of the partonic process. In order to ensure $x_A$ to be very small, we will tag the final state particles within the FoCal acceptance, i.e.\ both $k_1$ and $k_2$ will have large rapidity, so that $x_A\ll x_B$ due to relations
\begin{equation}
    x_A = \sum_i \frac{|\vec{k}_{T_i}|}{\sqrt{s}} e^{-\eta_i} ,\quad  x_B = \sum_i \frac{|\vec{k}_{T_i}|}{\sqrt{s}} e^{ \eta_i} \,,
\end{equation}
where the sum goes over final states.
We will assume that $x_A$ is small enough to apply the saturation formalism, i.e.\ nonlinear evolution equations for the $A$ target, whilst $x_B$ is moderate, so that no nonlinear evolution is needed there (we shall explicitly demonstrate this asymmetry  in our computation in Section~\ref{sec:results}). Such a concept is known as the hybrid factorization \cite{Dumitru:2005gt,Deak:2009xt}. Consequently, to a good approximation, only gluons are exchanged between the hard process and the target $A$.

In order to describe the target $A$ we shall use the so-called Transverse Momentum Dependent (TMD) gluon distributions. There are at least two reasons for that.
The first one is related to the observables we want to study. Having two final states allows us to investigate azimuthal plane correlations between the final states, which is directly related to the transverse momenta of partons.
The second reason is that we shall use the high energy QCD, where the concept of transverse momentum dependent gluon distributions is rather fundamental. This is because in the high energy regime, or, equivalently the small-$x$ limit, the largest scale is set by $s\rightarrow \infty$ at fixed $Q^2$, where $s$ is CM energy squared and $Q^2$ is the hard scale (given for example by the transverse momenta of jets). In collinear factorization, on the other hand, we have $Q^2\rightarrow \infty$ at fixed $s$. Therefore, at small $x$ one has to include the power corrections $k_T/Q$, which are not necessarily suppressed. 
This means one should not neglect the transverse momenta of exchanged gluons, both in the PDF, which then becomes the TMD PDF, and in the hard process, which in that case has to be computed from off-shell amplitudes. This framework is known as the High Energy Factorization (HEF) or $k_T$-factorization \cite{Catani:1990eg,Catani:1994sq}. It is also possible (and essential in the context of the present work) to incorporate a nonlinear evolution and thus the saturation phenomenon in the TMD PDFs.

In QCD the TMD PDFs have the following generic operator definition
\begin{equation}
\mathcal{F}\left(x,k_T\right) = 2\int\frac{d\xi^{-}d^{2}{\xi_T}}{(2\pi)^{3}P^{+}}\, e^{\,ix P^{+}\xi^{-}-i\vec{k}_{T}\cdot\vec{\xi}_{T}} 
\left\langle P\right|
\mathrm{Tr} \hat{F}^{j+} (\xi^-,\vec{\xi}_T,0 )\mathcal{U}_1
\hat{F}^{j+}\left(0\right)\mathcal{U}_2
\left|P\right\rangle ,
\label{eq:TMD_generic}
\end{equation}
where $|P\rangle$ is a hadron state with momentum $P$, $\hat{F}^{j+}=F^{j+}_at^a$ is the gluon field strength tensor projected onto a transverse component $j$ and light-cone ``plus'' component. The two field operators are displaced in the light cone ``minus'' direction by an amount $\xi^-$ and in the transverse direction by a vector $\vec{\xi}_T$. For simplicity, the transverse components are summed over, what corresponds to the unpolarized TMD PDF. $\mathcal{U}_1$ and $\mathcal{U}_2$ are Wilson lines in fundamental representations connecting the points where the fields are evaluated. (In principle, there could be a double-trace expression in the definition, but we omit here this case for brevity.) The precise form of these Wilson lines is, in principle, arbitrary; it is not fixed by the requirement of gauge invariance. However, resummation of gluons collinear to the target hadron predicts the form of these Wilson lines and their structure depends on the color flow in the hard process \cite{Bomhof:2006dp}. For a discussion of all possible operators (following the prescription of \cite{Bomhof:2006dp}) and explicit calculation of operators relevant for some multiparticle processes see \cite{Bury:2018kvg}. 

Consider now the simplest partonic sub-process contributing to \eqref{eq:main_process}, that dominates at small $x$:
\begin{equation}
    g(k_A)+q(k_B) \rightarrow \gamma(k_1) + q(k_2) \,.
\end{equation}
The corresponding TMD gluon PDF associated with the gluon $k_A$ (whose longitudinal fraction $x_A$ is small) has a one-term from which follows a very simple color structure of that sub-process. It reads
\begin{equation}
\mathcal{F}_{qg}^{(1)}\left(x,k_T\right) = 2\int\frac{d\xi^{-}d^{2}{\xi_T}}{(2\pi)^{3}P^{+}}\, e^{\,ix P^{+}\xi^{-}-i\vec{k}_{T}\cdot\vec{\xi}_{T}} 
\left\langle P\right|
\mathrm{Tr} \hat{F}^{j+} (\xi^-,\vec{\xi}_T,0 )\mathcal{U}^{[+]}
\hat{F}^{j+}\left(0\right)\mathcal{U}^{[-]}
\left|P\right\rangle ,
\label{eq:TMD_dipole}
\end{equation}
where $\mathcal{U}^{[-]}$ and $\mathcal{U}^{[+]}$ are, respectively, past-pointing and future-pointing ``staple-like" Wilson lines:
\begin{equation}
    \mathcal{U}^{[\mp]} =\! \left[(\xi^-,\vec{\xi}_T,0),(\mp\infty,\vec{\xi}_T,0)\right]\!
    \left[(\mp\infty,\vec{\xi}_T,0),(\mp\infty,\vec{0}_T,0)\right]\!
    \left[(\mp\infty,\vec{0}_T,0),(0,\vec{0}_T,0)\right] \, ,
    \label{eq:U-}
\end{equation}
where $\left[x,y\right]$ are straight Wilson line segments, for example
\begin{equation}
    \left[(\xi^-,\vec{0}_T,0),(0,\vec{0}_T,0)\right] = \mathcal{P} \exp\left\{ig \int_{0}^{\xi^-}\!\!
    ds \hat{A}^+(s,\vec{0}_T,0)\right\} \, .
    \label{eq:StraightWilson}
\end{equation}
The TMD gluon distribution in \eqref{eq:TMD_dipole} cannot be interpreted as the gluon number distribution; it is not possible to choose the gauge that would eliminate both the past-pointing and the future-pointing Wilson lines. 
Such situation would be the case if one considered instead the production of dijets in DIS, where there is only one colored initial state. 

In the present work we are interested in the small-$x$ limit of the TMD \eqref{eq:TMD_dipole}. It has been recognized in \cite{Dominguez:2011wm} that \eqref{eq:TMD_dipole} corresponds to the so-called dipole gluon distribution known in the CGC theory (see also  \cite{Kharzeev:2003wz} for the discussion of two different gluon distributions at small $x$). This correspondence built a bridge (or rather a footbridge at that time) between the TMD factorization approach and the CGC. More precisely, one finds \cite{Dominguez:2011wm} that the CGC expressions for various  massless two-particle production processes correspond in the back-to-back limit to a convolution of a leading-twist TMD given by a generic formula \eqref{eq:TMD_generic} and an on-shell hard factors, corresponding to different color structures entering the TMD. 

On the other hand, two-particle correlations are extremely interesting beyond the back-to-back limit, when the two-particle imbalance is treated more precisely. This can be taken into account via the so-called small-$x$ Improved TMD (ITMD) factorization \cite{Kotko:2015ura}. This formalism includes the power corrections (i.e.\ the $k_T$ dependence) in the hard factors, which are computed within the high energy factorization framework \cite{Catani:1990eg}, using the modern automated methods \cite{vanHameren:2012uj,vanHameren:2012if,Kotko:2014aba}, all consistent with the Lipatov's effective action \cite{Lipatov:1995pn}. In \cite{Altinoluk:2019fui} it was demonstrated, that the ITMD formulation can be rigorously derived from the CGC, for certain processes, by neglecting the so-called geniune twist corrections, i.e.\ higher twist operators that are normally associated with ``hard'' multiple partonic interactions. These type of multiple interaction contributions are suppressed for two-body final states with the hard scale and thus for processes like jet production the ITMD formalism is very convenient \cite{vanHameren:2016ftb,Fujii:2020bkl,Boussarie:2021ybe,Altinoluk:2021ygv}.

In particular, in \cite{Altinoluk:2019fui} the ITMD factorization formula for the photon+jet production was explicitly derived, confirming that in that case, there are no higher genuine twist contributions, thus the formula essentially reproduces earlier CGC result of \cite{Dominguez:2011wm}, and the corresponding operator is the one in \eqref{eq:TMD_dipole}. There is a formal difference between CGC and ITMD though. In ITMD the operator definitions of gluon distribution functions are more general, to leading genuine twist, as they incorporate full $x$ dependence (for a related discussion see \cite{Boussarie:2020fpb}). The LO factorization formula reads
\begin{equation}
    d\sigma_{pA\rightarrow j+\gamma} = \int\! dx_A\, dx_B \int\! d^2k_T\, f_{q/B}(x_B;\mu)\,\mathcal{F}_{qg}^{(1)}\left(x_A,k_T;\mu \right)   d\sigma_{qg^*\rightarrow q\gamma}(x_A,x_B,k_T;\mu) \,,
    \label{eq:ITMD_factoriz}
\end{equation}
where $f_{q/B}$ is the collinear PDF for a quark in target $B$ (a proton), $\mathcal{F}_{qg}^{(1)}$ is the dipole TMD PDF \eqref{eq:TMD_dipole} corresponding to target $A$ (proton or nucleus) and $d\sigma_{qg^*\rightarrow q\gamma}$ is the LO partonic cross section obtained from the amplitude with off-shell initial state gluon $g^*(k_A)+q(k_B) \rightarrow \gamma(k_1) + q(k_2)$, where the incoming momenta are
\begin{gather}
    k_A^{\mu}=x_A P_A^{\mu} + k_T^{\mu}\,, \\
    k_B^{\mu}=x_B P_B^{\mu} \,,
\end{gather}    
with $k_T\cdot P_A=k_T\cdot P_B = 0$. In the high energy kinematics the off-shell gluon couples eikonally, i.e.\ through $p_A^{\mu}$ instead of the transverse polarization vector. Obviously, since the TMD PDF is gauge invariant, the amplitude has to defined in a gauge invariant way as well. This is, in general, non trivial with off-shell gluons and requires introducing Wilson line along $P_A$, see eg.~\cite{Kotko:2014aba}. However, in the present simple process gauge invariance is satisfied trivially off-shell with the ordinary set of diagrams. 

Notice, that in the factorization formula \eqref{eq:ITMD_factoriz} we have introduced the dependence of the TMD PDF on the hard scale $\mu$. This is an effective (and efficient) way of incorporating the resummation of the Sudakov logarithms. They are important for observables sensitive to $k_T$, like azimuthal correlations, when a hard scale $\mu\gg k_T$ is present. At small $x$, in particular in the CGC formalism, the Sudakov resummation was carried out for various processes \cite{Mueller:2012uf,Mueller:2013wwa,Zheng:2014vka,Mueller:2015ael,Stasto:2018rci,Caucal:2022ulg,Taels:2022tza,Caucal:2023nci} as well as the threshold resummation and combination of both \cite{Xiao:2018zxf,Liu:2020mpy,Liu:2022ijp}.
The resummation is done in the impact parameter space and, in general, entangles the hard part, the collinear PDFs and the TMD PDFs. 
On the other hand, there are many studies of small-$x$ TMD gluon distributions with explicit hard scale dependence, for example \cite{Balitsky:2015qba,Balitsky:2016dgz}. Further, more relevant phenomenologically, there were attempts in the linear domain to merge the BFKL and DGLAP equations \cite{Kwiecinski:1997ee}, \cite{Kimber:1999xc,Kimber:2001sc}, or to describe the evolution in terms of the coherent gluon emission  \cite{Catani:1989sg,Catani:1989yc}. In the nonlinear domain, the notable work is an attempt \cite{Kutak:2011fu} to extend the CCFM equation to account for nonlinear term. Finally, in \cite{vanHameren:2014ala,Kutak:2014wga} an effective methods of implementing the hard scale dependence was proposed, based on a reweighting procedure which can be applied both to linear and non-linear evolution.
Although, in perturbative QCD, in general  the Sudakov factor cannot be distributed to provide the same hard scale dependence in the collinear PDF and TMD PDF as in  \eqref{eq:ITMD_factoriz}, in \cite{Al-Mashad:2022zbq} the Authors compared the full impact parameter resummation with the simplified approach for dijet production in pA collisions, and they are qualitatively very similar. We expect similar thing to occur in the present case, therefore we shall proceed with the formula \eqref{eq:ITMD_factoriz}.

\section{Results}
\label{sec:results}
In this section we present numerical results for photon+jet production in p-p and p-Pb collisions at 8.8~TeV, within the FoCal acceptance range. In order to allow for jet reconstruction we choose the pseudorapidity range to be $3.8< \eta < 5.1$. We study jets and photons with the minimum transverse momentum ranging between 5-20 GeV.
The CM energy of 8.8~TeV per nucleon is the declared value for which there will be both p-p and p-Pb measurement. In addition, we compute p-p cross sections for 5 and 14 TeV, in order to see how the gluon saturation pattern in protons changes with energy.

The calculations were performed within the ITMD formalism, Eq.~\eqref{eq:ITMD_factoriz}, which at tree level is implemented in the parton-level Monte Carlo generator \KaTie~\cite{vanHameren:2016kkz}. Using the Monte Carlo allows for convenient application of the kinematic cuts and study of multiple observables. The collinear PDFs were provided by the LHAPDF library \cite{Buckley_2015} and set to CT18NLO. The dipole TMD PDF with hard scale dependence is based on the Kutak-Sapeta (KS) fit of the nonlinear evolution equation of \cite{Kutak:2003bd,Kwiecinski:1997ee} to HERA data \cite{Kutak:2012rf} and were obtained in \cite{Al-Mashad:2022zbq}. We require the minimal distance in the $\eta-\phi$ plane, where $\phi$ is the azimuthal angle, of $R=0.4$ between the photon and the jet. Both the factorization and renormalization scales are set to the average $p_T$, $\mu_F = \mu_R = (p_{T_1} + p_{T_2})/2$.

\begin{figure}
     \centering
     \begin{subfigure}[h]{\textwidth}
         \centering
         \includegraphics[width=8cm]{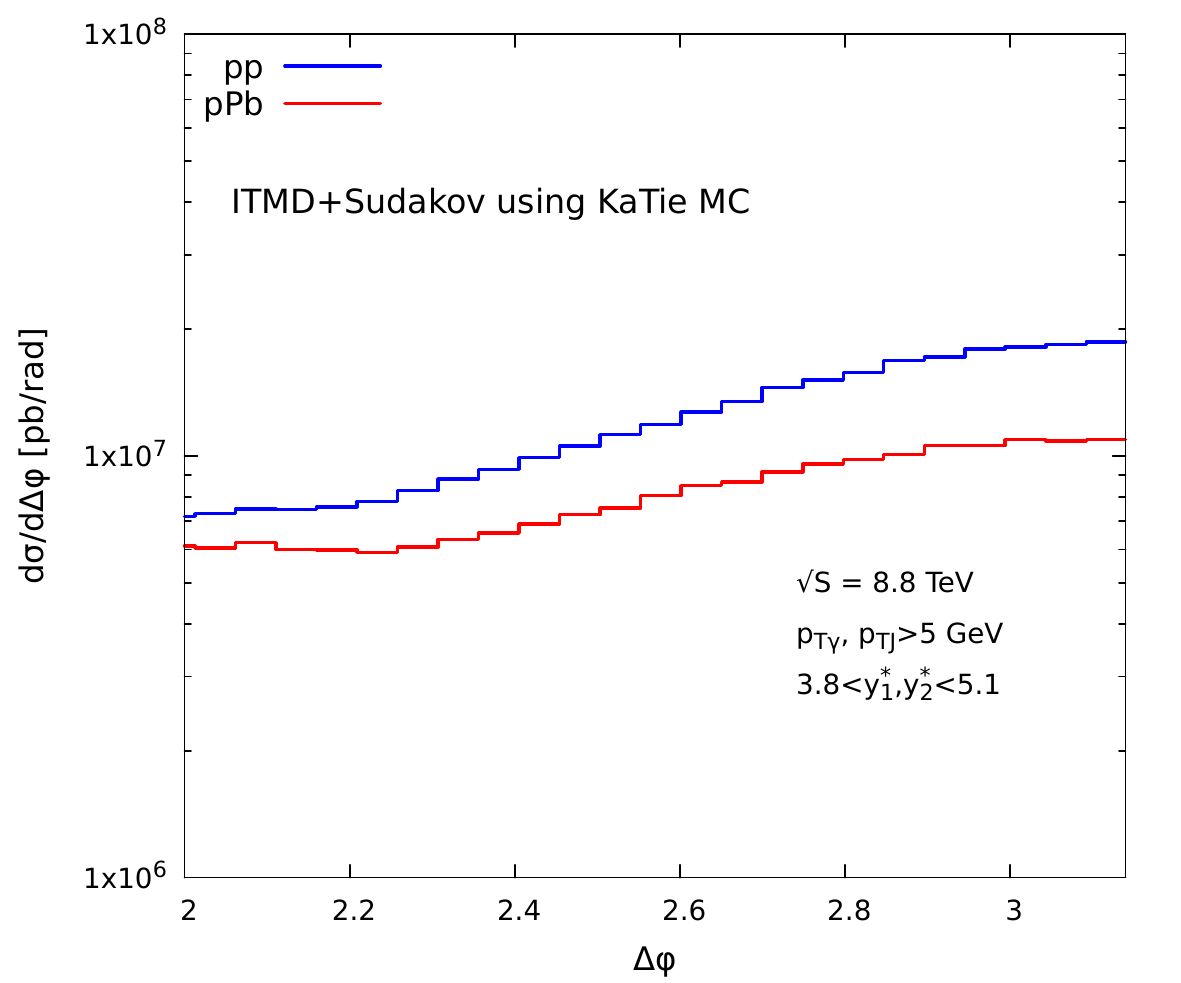}
         \label{dphi_pT5}
     \end{subfigure}\\
     \begin{subfigure}[h]{\textwidth}
         \centering
         \includegraphics[width=8cm]{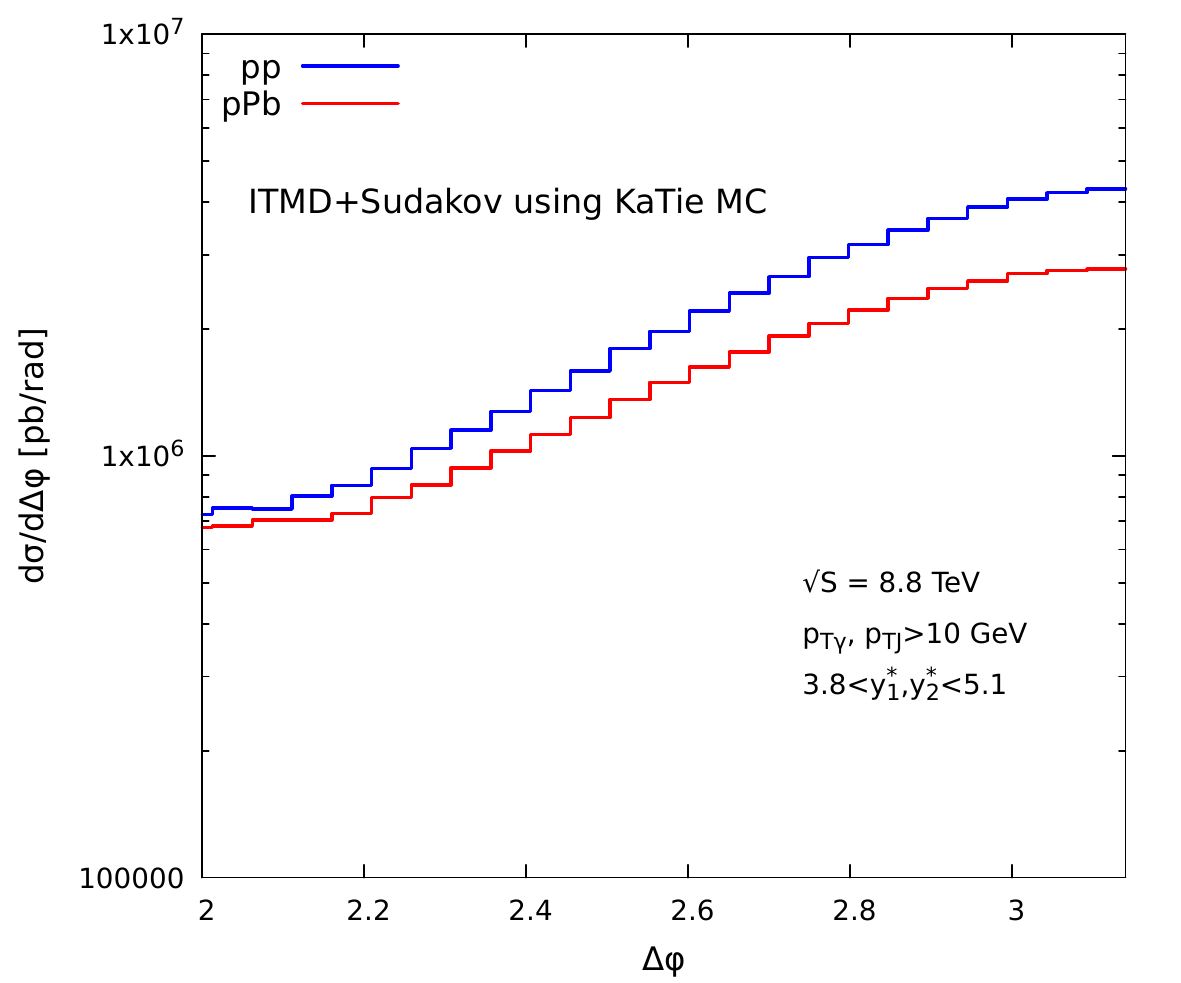}
         \label{dphi_pT10}
     \end{subfigure} \\
     \begin{subfigure}[h]{\textwidth}
         \centering
         \includegraphics[width=8cm]{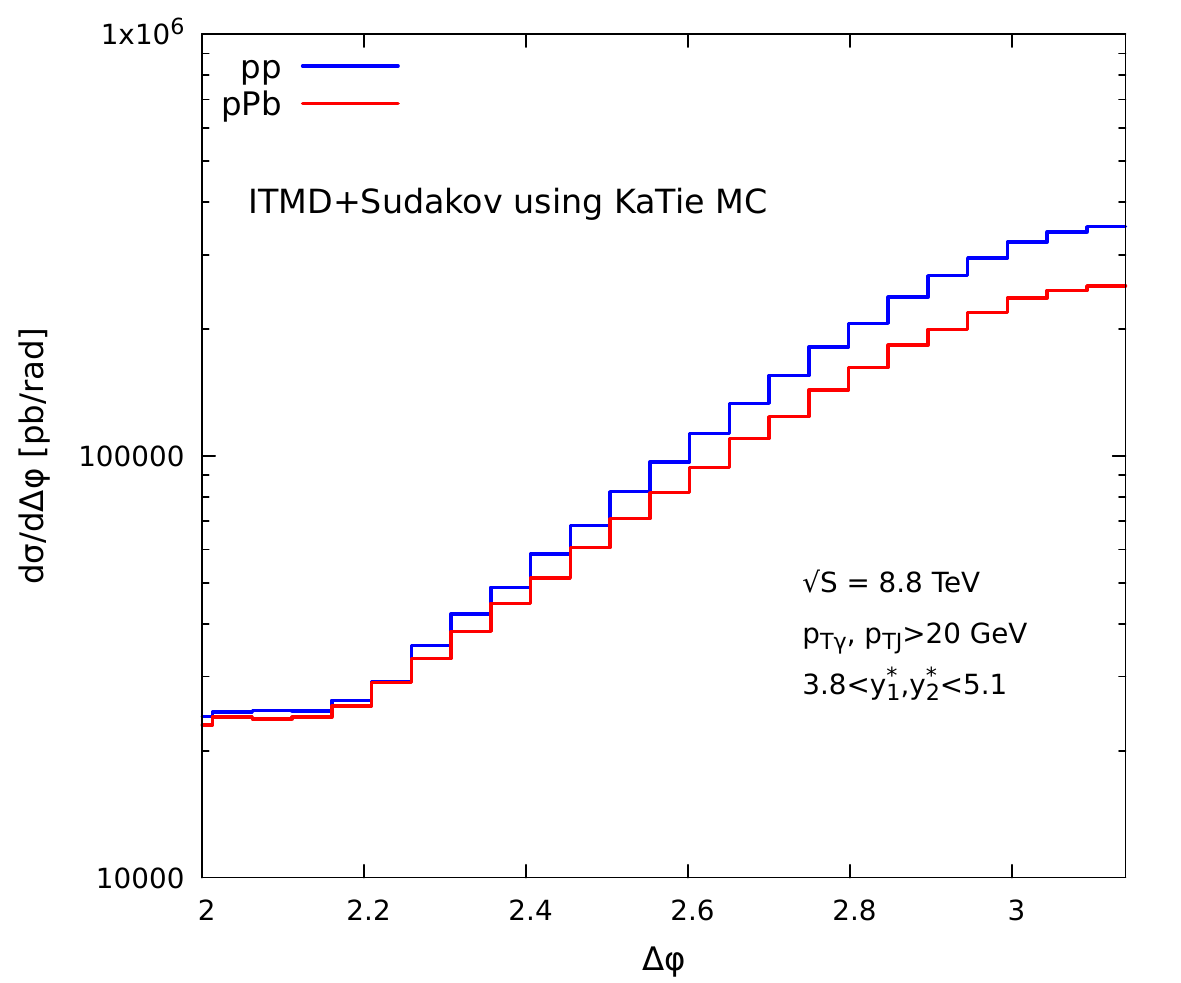}
         \label{dphi_pT20}
     \end{subfigure}
    \caption{
    Differential cross sections for $\gamma$+jet in FoCal acceptance range for p-p and p-Pb collisions at $\sqrt{s} = 8.8$TeV, as a function of azimuthal separation $(\Delta \varphi)$. The three plots correspond to different transverse momentum cuts on final states: 5~GeV, 10~GeV and 20~GeV.
    }
    \label{dphi}
\end{figure}

\begin{figure}
    \centering
    \includegraphics[width=8cm]{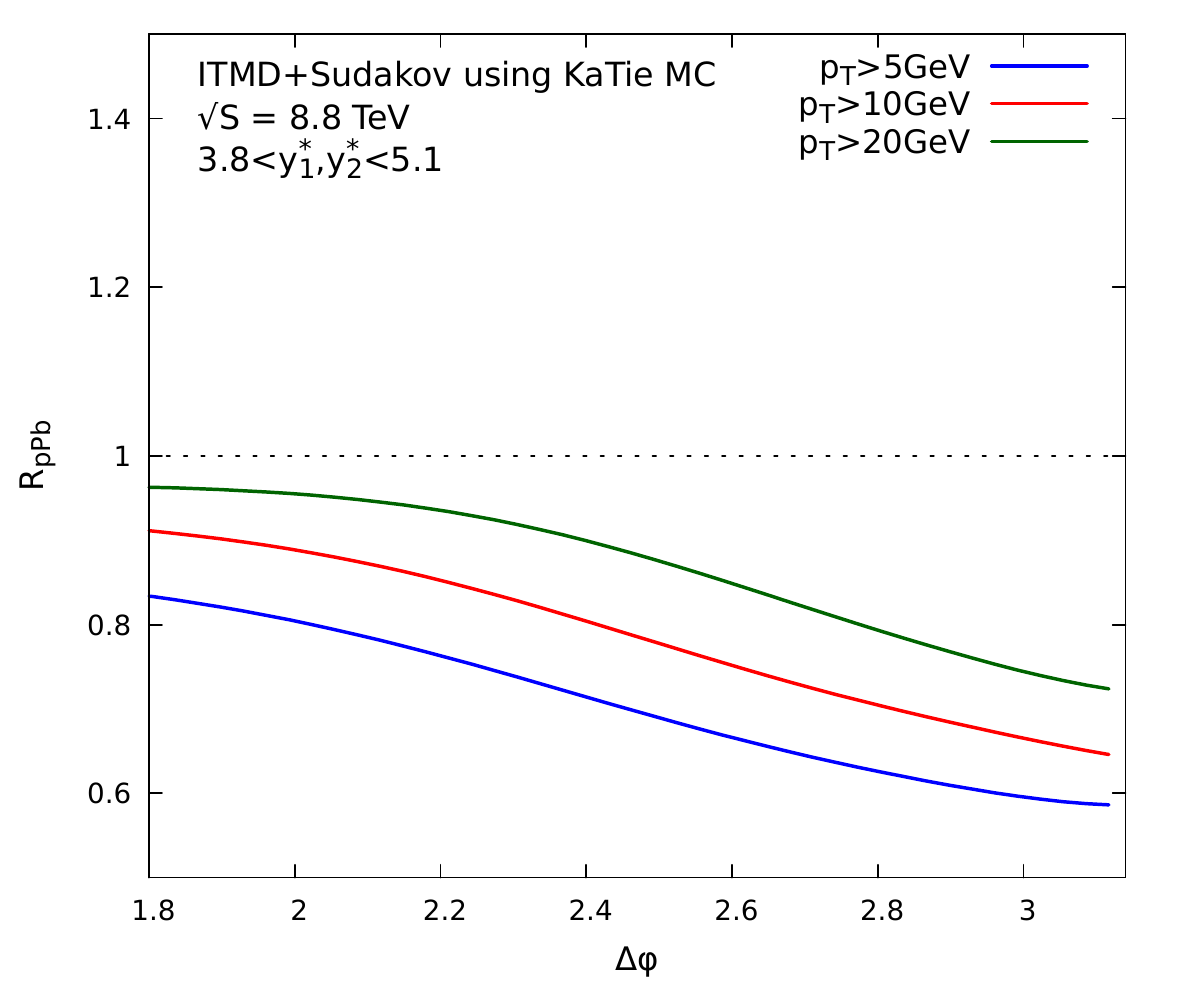}
    \caption{Nuclear modification ratio as a function of the azimuthal angle difference $\Delta \varphi$ for $\gamma$ + jet with three different $p_T$ thresholds.}
    \label{RpA}
\end{figure}

The first observable we studied are the azimuthal correlations. In Fig.~\ref{dphi} we show the differential cross section as a function of the azimuthal angle between the photon and the jet for p-p and p-Pb collisions, $\textrm{d} \sigma / \textrm{d} \Delta \phi$, for three different cuts on the transverse momenta, 5~GeV, 10~GeV and 20~GeV. Notice, the p-Pb cross section is divided by the number of nucleons. We observe a clear and large suppression of the back-to-back peak of p-Pb comparing to p-p indicating very strong saturation signal. The suppression gets milder with increasing the $p_T$ cut. It is essential to notice, that  including the Sudakov resummation the saturation signal, understood as the suppression of p-Pb with respect to p-p, is not washed away. We summarize this fact in Fig.~\ref{RpA} by computing the nuclear modification ratios
\begin{equation}
    \textrm{R}_{\textrm{pPb}} = \frac{\big(\frac{d \sigma } { d \Delta \phi}\big)_{\mathrm{pPb}}}{\big(\frac{d \sigma } { d \Delta \phi}\big)_{\mathrm{pp}}} \,.
\end{equation}
We see that the suppression of the p-Pb cross section due to saturation is up to 40\% for the lowest $p_T$ cut, and decreases to about 20\% for $p_T>20\,\mathrm{GeV}$. Similar conclusions were obtained for di-jets in \cite{Al-Mashad:2022zbq}.


\begin{figure}
     \centering
         \includegraphics[width=8cm]{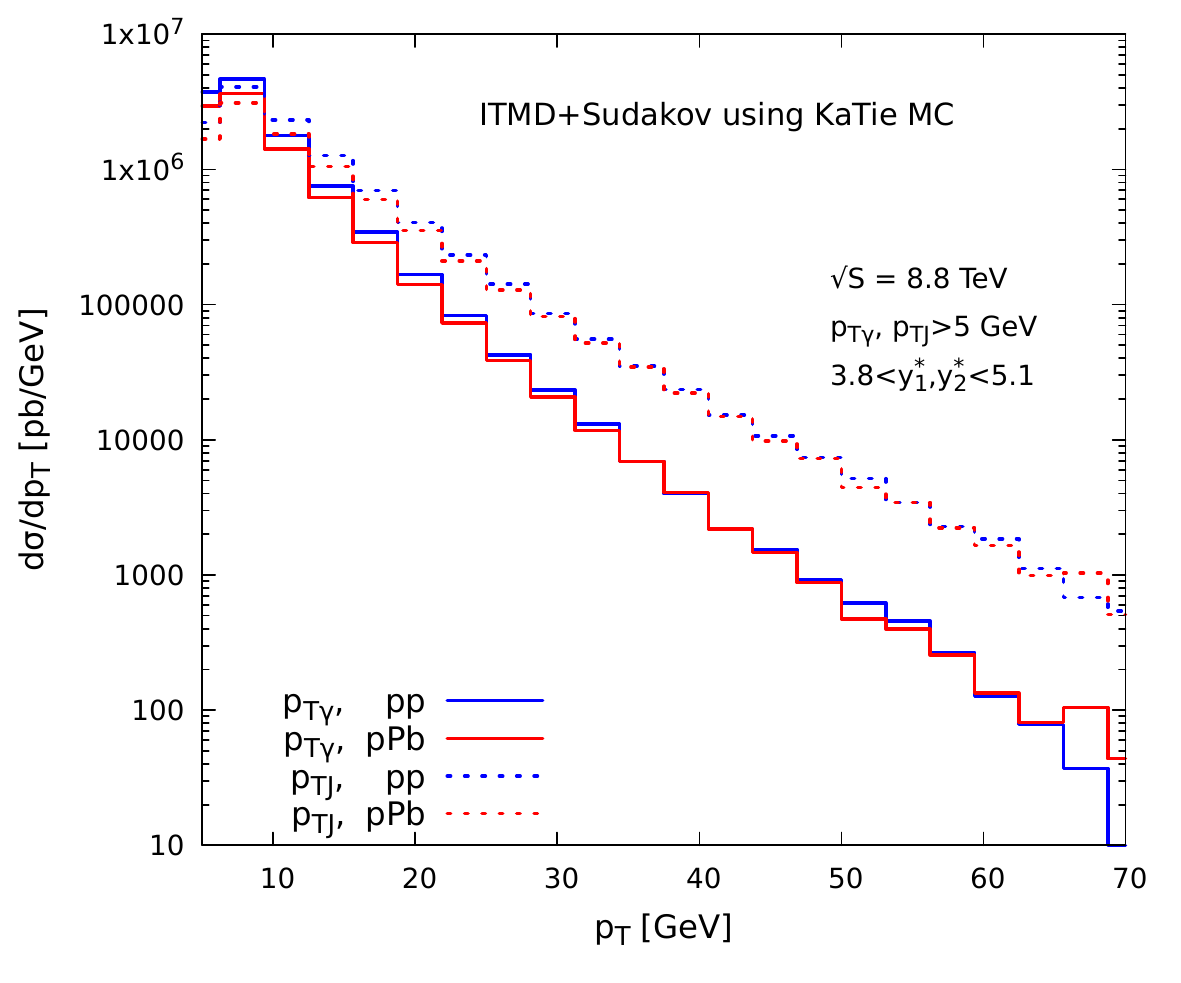}
    \caption{Transverse momentum distributions for photons and jets in FoCal acceptance range for p-p and p-Pb collisions at $\sqrt{s} = 8.8$~TeV. }
    \label{pt}
\end{figure}

Next, in Fig.~\ref{pt} we show the transverse momentum spectra of photons and jets for p-p and p-Pb collisions. We observe that the photon spectrum is softer than jet spectrum. One should however remember that transverse momentum spectra are in general more affected by final state showers and hadronization effects, which are not included in the present computation.


\begin{figure}
    \centering
    \begin{subfigure}[h]{0.49\textwidth}
    \includegraphics[width = \textwidth, height = 6cm]{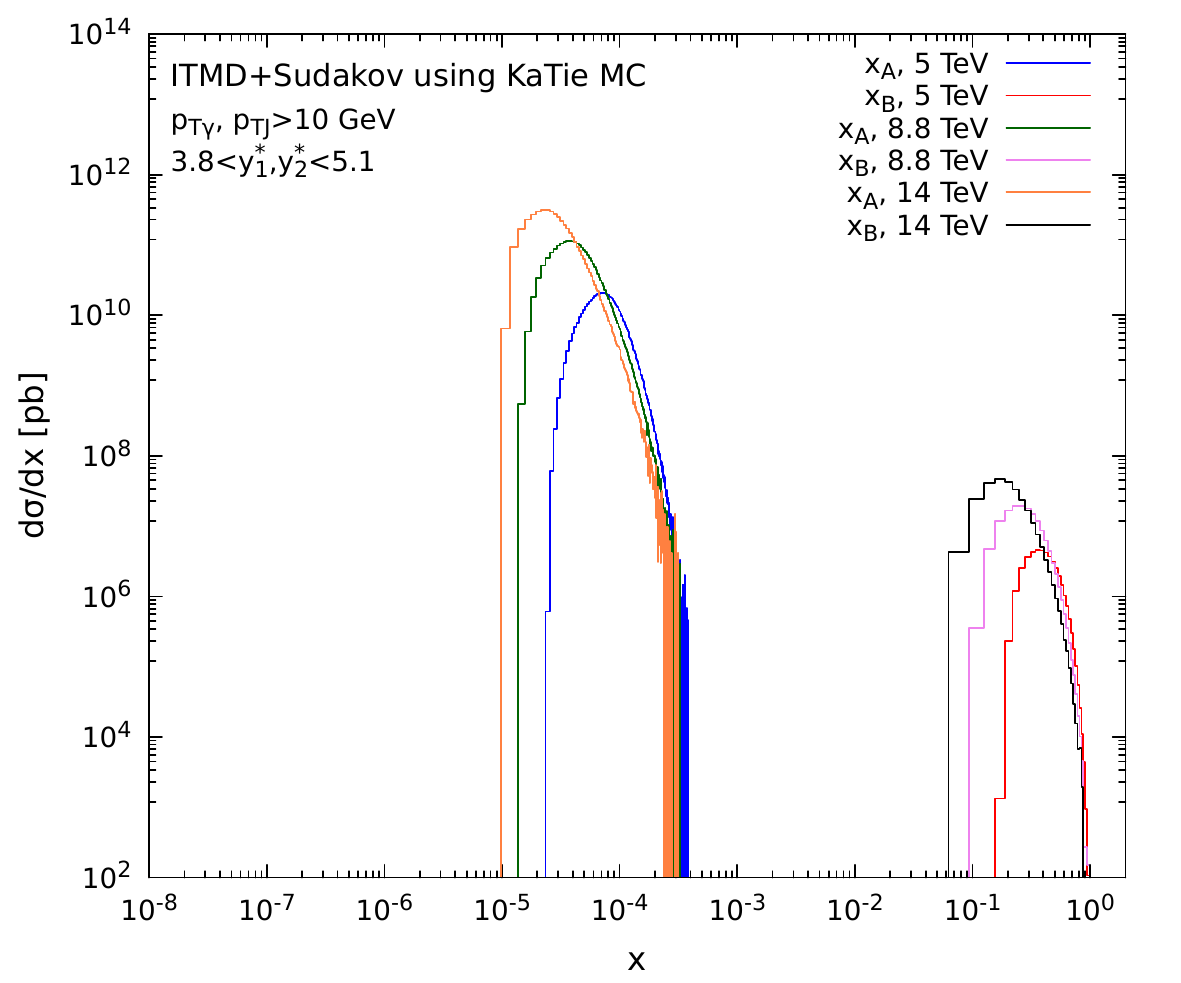}
    \caption{
    \small
    Spectra of longitudinal fractions probed in the TMD PDF, $x_A$, and the collinear PDF, $x_B$ for $\gamma$+jet production at different $\sqrt{s}$ energies in FoCal acceptance for moderate transverse momentum cut $p_T>10\,\mathrm{GeV}$.}
    \label{xplot1}
    \end{subfigure}
    \hfill
\begin{subfigure}[h]{0.49\textwidth}
    \centering
    \includegraphics[width = \textwidth, height = 6cm]{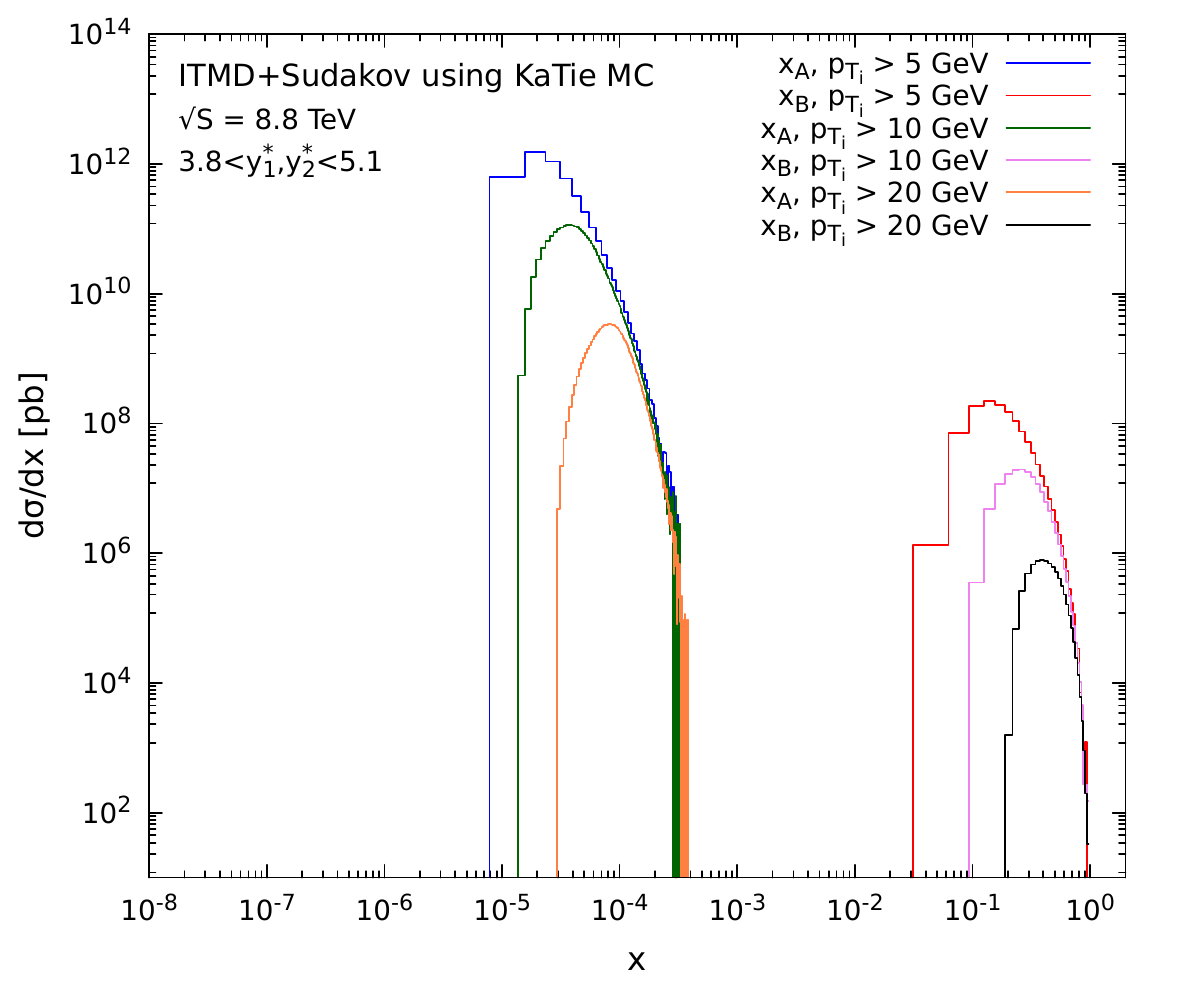}
    \caption{
    \small
    \label{xplot2}Spectra of longitudinal fractions probed in the TMD PDF, $x_A$, and the collinear PDF, $x_B$ for $\gamma$+jet production at $\sqrt{s}$ energy 8.8 TeV in FoCal acceptance for different transverse momentum cut $p_T>5\,\mathrm{GeV}$, $p_T>10\,\mathrm{GeV}$ and $p_T>20\,\mathrm{GeV}$.}
    
\end{subfigure}
\end{figure}

In Fig.~\ref{xplot1} we show the differential cross sections in $x_A$ and $x_B$ for the moderate $p_T$ cut 10~GeV. We clearly see that we do fulfill the asymmetry requirement $x_A \ll x_B$ enabling us to use the hybrid $k_T$ factorization. Moreover, the $x_A$ which is probed is typically smaller then 10$^{-4}$, which further justifies usage of the small $x$ formalism. The collinear partons are typically probed at quite large $x_B\sim 0.2$. The present computation could therefore in principle be improved by including the threshold resummation. However, the present range of $x_B$ is still reasonable to treat with the pure collinear framework.

  
\begin{figure}
    \centering
     \begin{subfigure}[h]{0.49\textwidth}
         \centering
         \includegraphics[height=5.5cm]{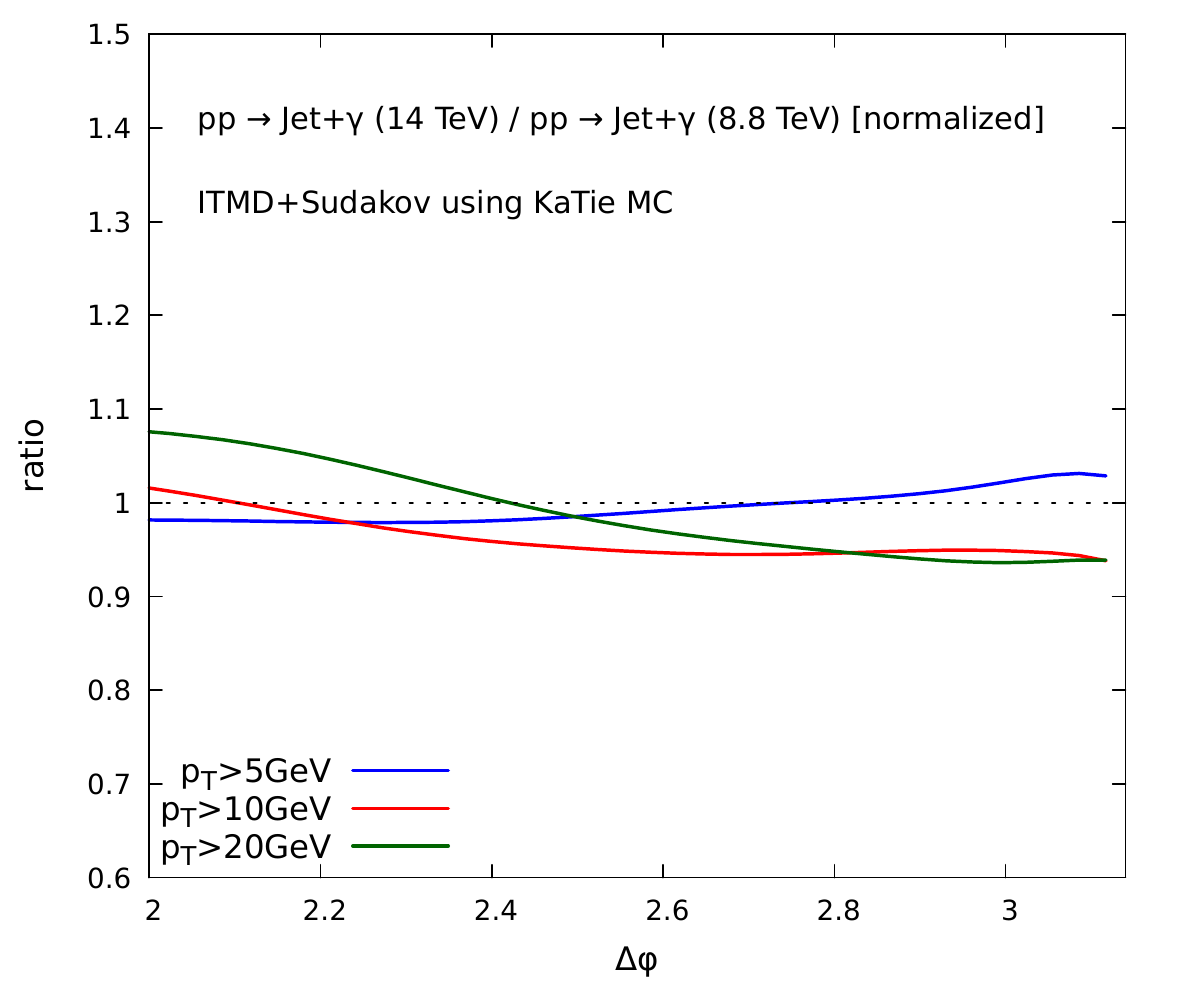}
    \end{subfigure}
    \begin{subfigure}[h]{0.49\textwidth}
         \centering
         \includegraphics[height=5.5cm]{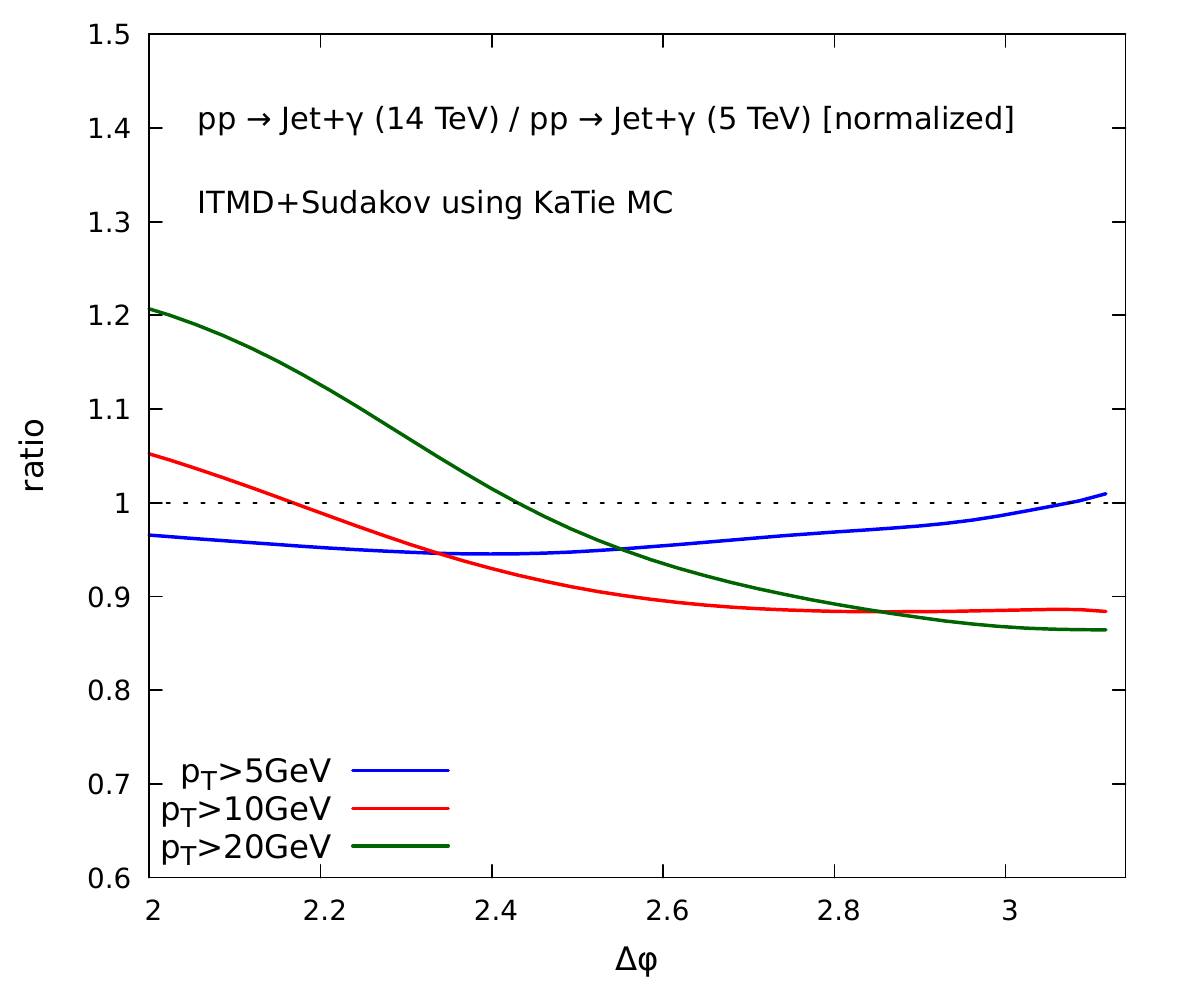}
    \end{subfigure}\\
     \begin{subfigure}[h]{0.49\textwidth}
         \centering
         \includegraphics[height=5.5cm]{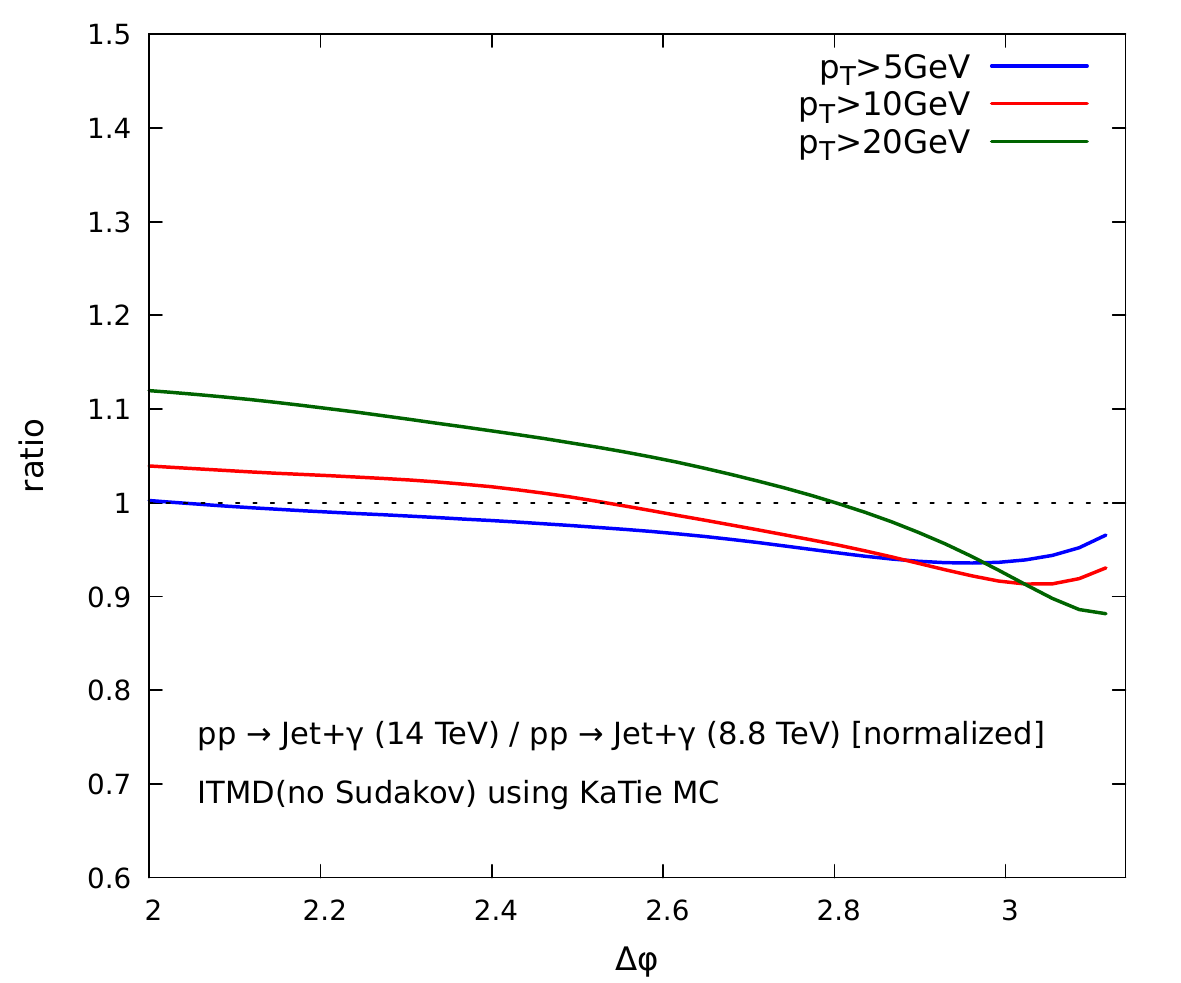}
    \end{subfigure}
    \begin{subfigure}[h]{0.49\textwidth}
         \centering
         \includegraphics[height=5.5cm]{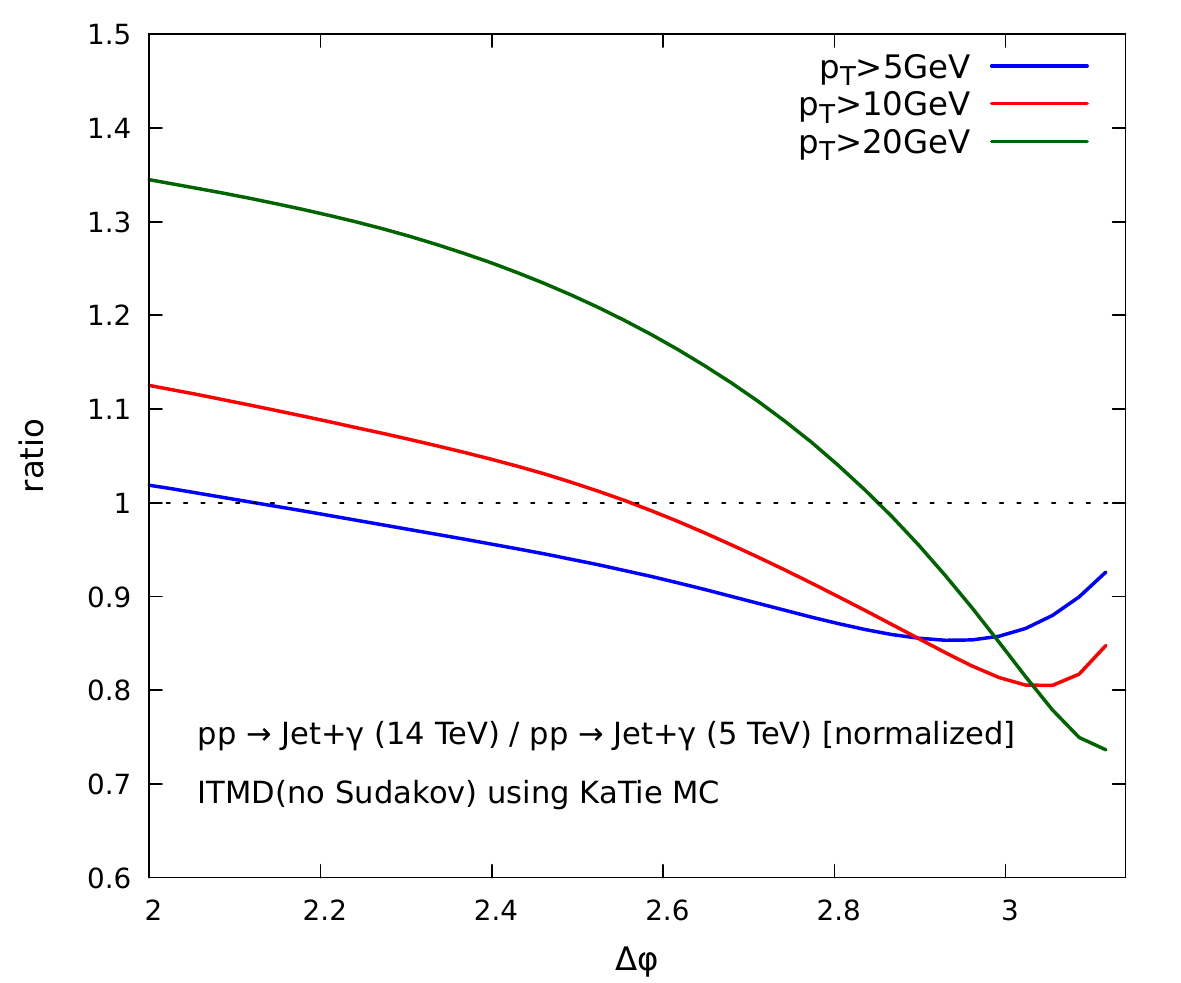}
    \end{subfigure}\\
     \begin{subfigure}[h]{0.49\textwidth}
         \centering
         \includegraphics[height=5.5cm]{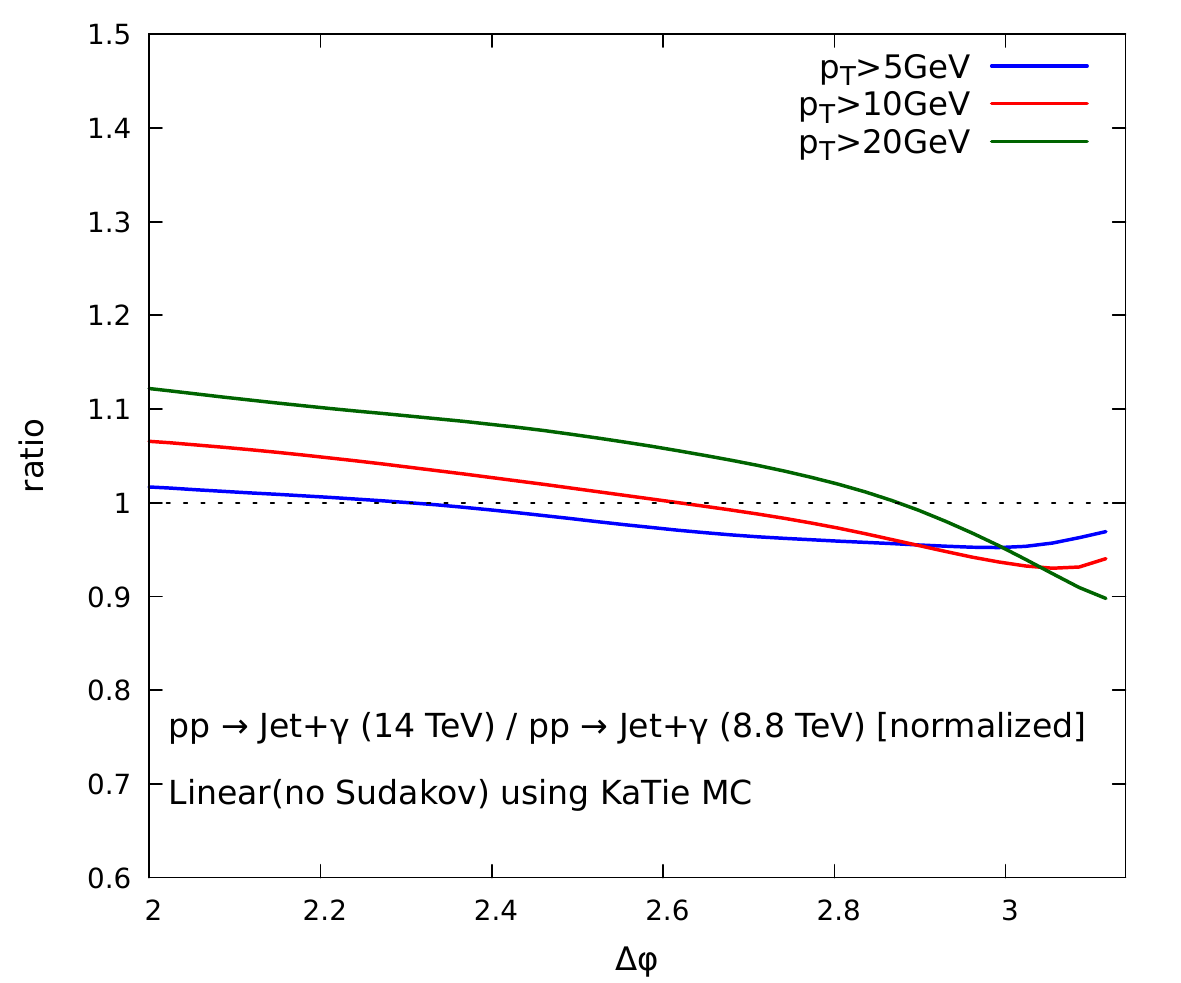}
    \end{subfigure}
    \begin{subfigure}[h]{0.49\textwidth}
         \centering
         \includegraphics[height=5.5cm]{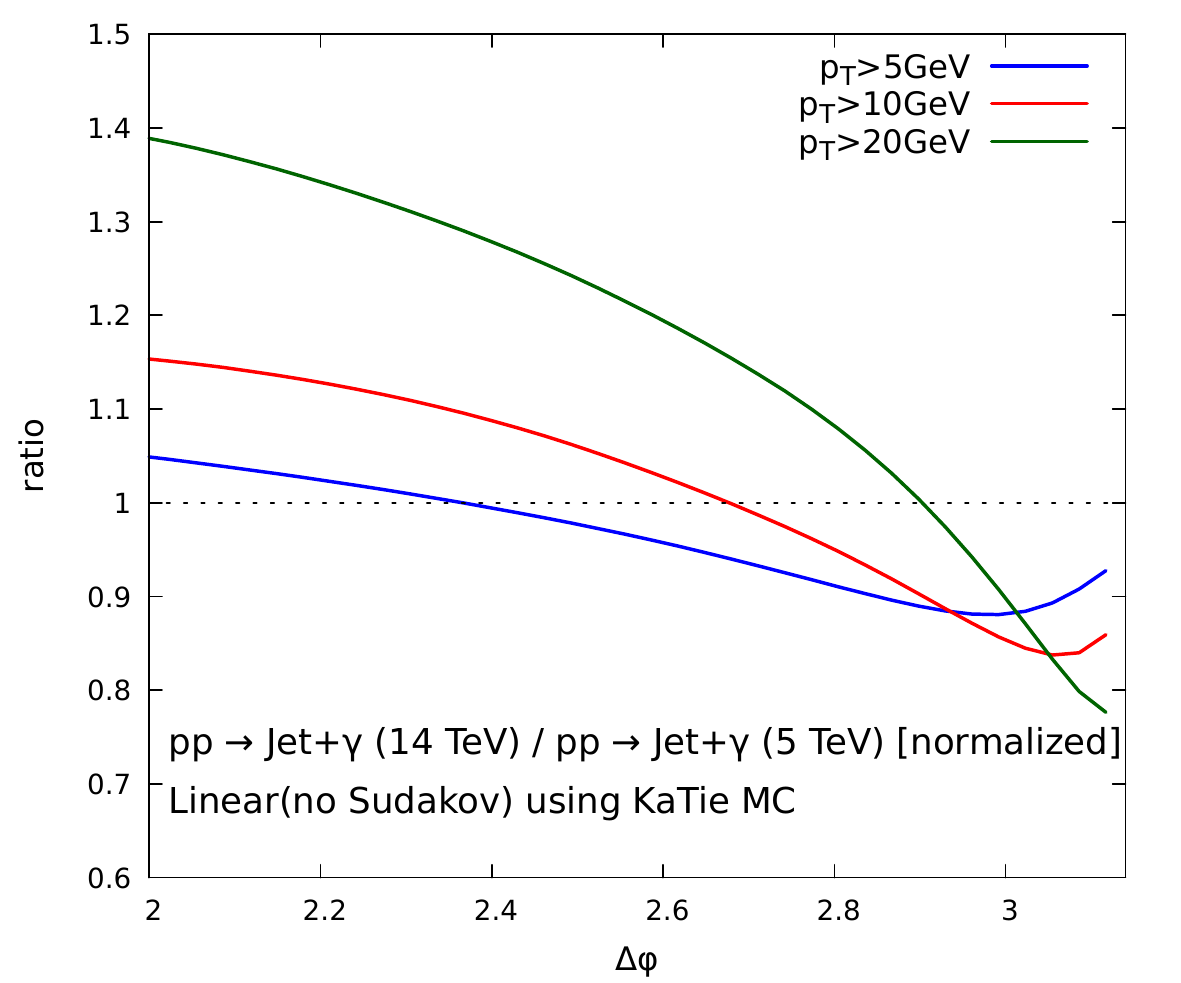}
    \end{subfigure}\\
    \begin{subfigure}[h]{0.49\textwidth}
         \centering
         \includegraphics[height=5.5cm]{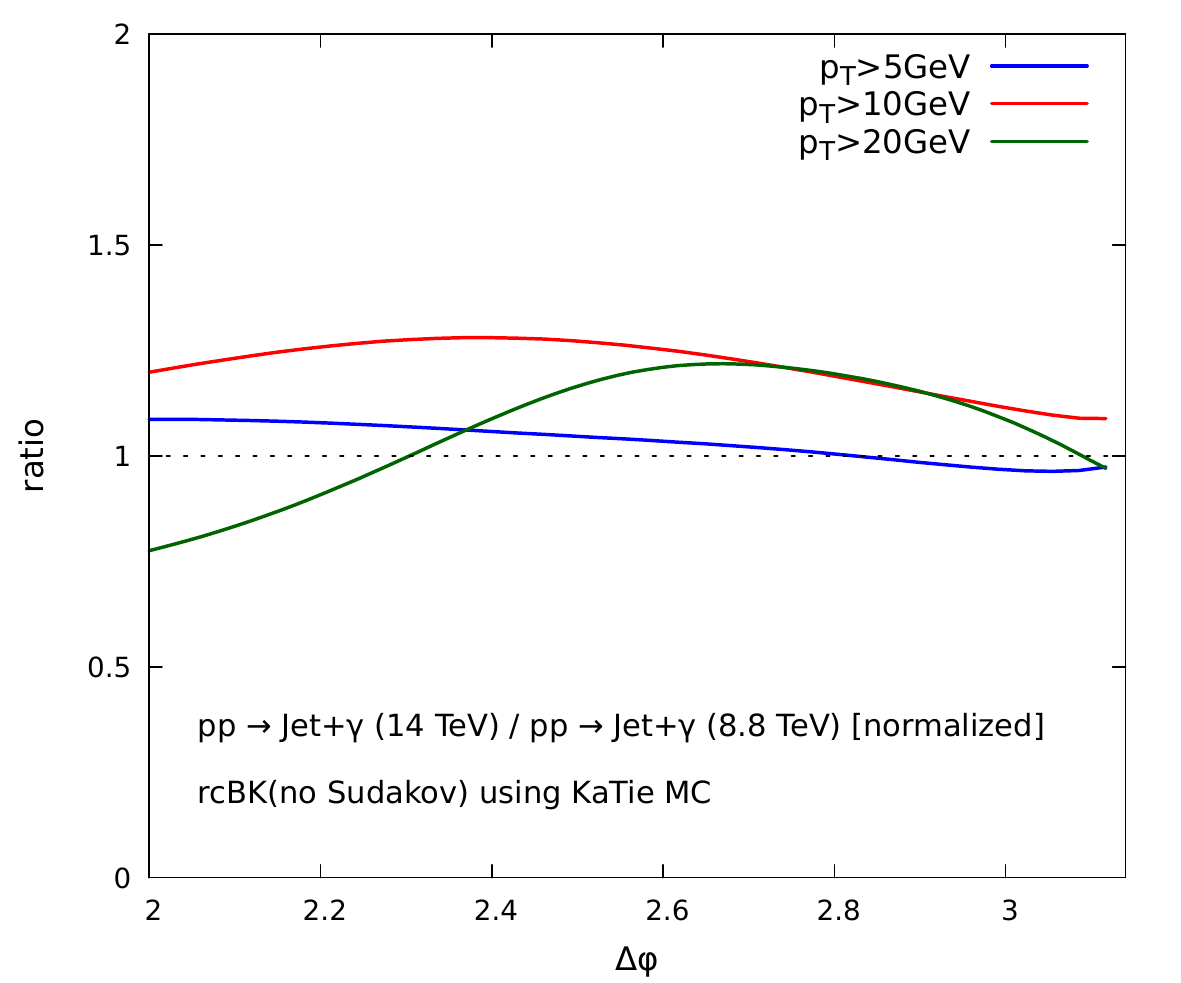}
    \end{subfigure}
    \begin{subfigure}[h]{0.49\textwidth}
         \centering
         \includegraphics[height=5.5cm]{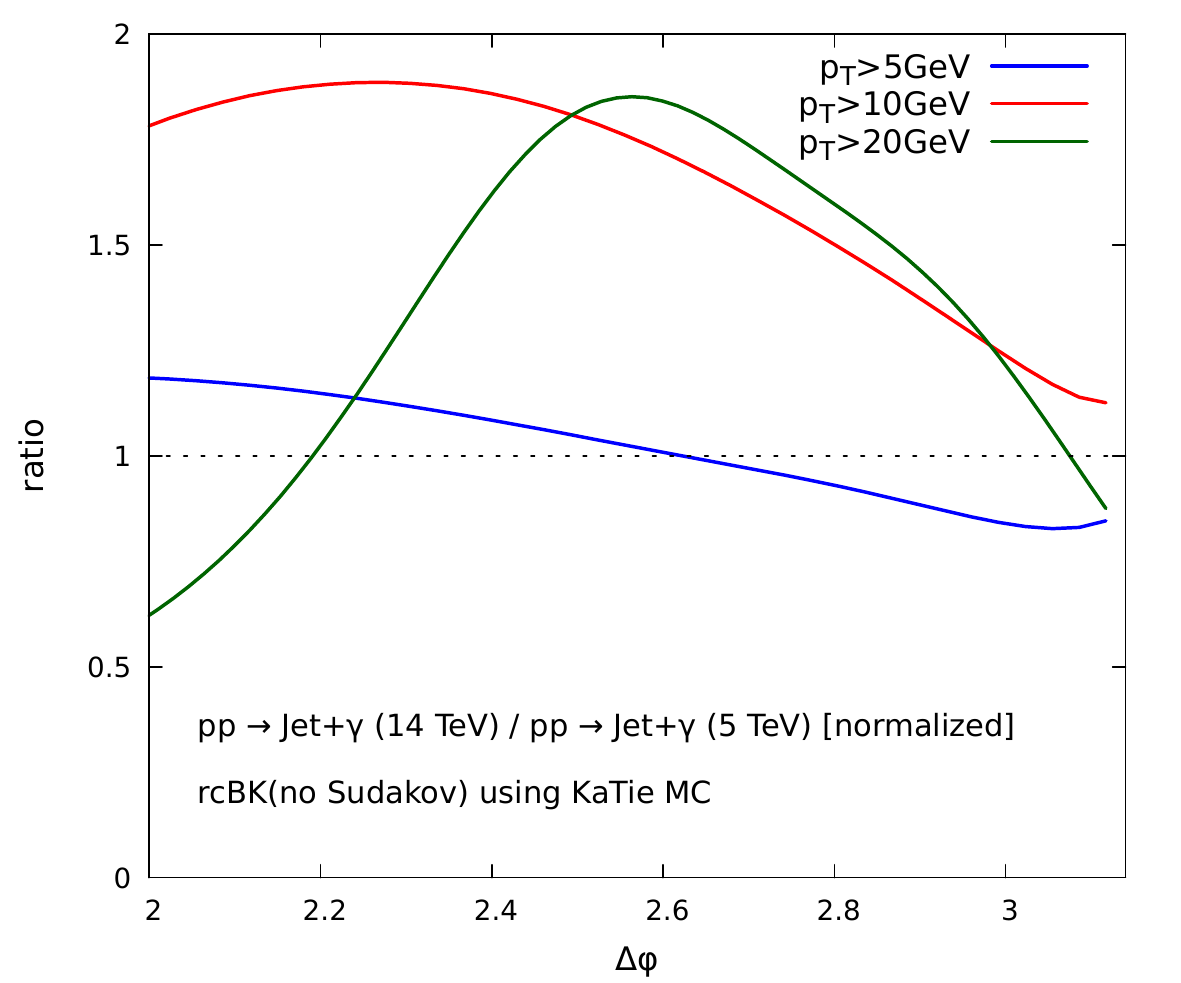}
    \end{subfigure}
\caption{\label{ratioplot1}
\small 
First row: Ratio of azimuthal angle distributions for $\gamma$+jet production in p-p collisions using ITMD+Sudakov for different energies and transverse momentum cuts. The left plot shows ratio of 14~TeV cross section to 8.8~TeV cross section, both independently normalized to the total cross section. The right plot shows similar ratio for 14~TeV cross section and 5~TeV.
Second row: Same as first row, but the calculations are done with TMD gluon distribution where the Sudakov resummation is not present.
Third row: The same as the second row, but the nonlinear term is turned off (and the resulting linear equation was fitted to HERA data). Fourth row: computation with the rcBK [Reference] dipole gluon distribution.
}
\end{figure}

Now we turn to discussion of the dependence of the p-p $\rightarrow \gamma$+jet cross section on the energy. Indeed, during FoCal operation it is planned to collect data at different energies, including. at least, 8.8 and 14.0~TeV. Since in the saturation formalism both nuclei and protons have to be treated within the non-linear evolution, one can ask what is the pattern of the p-p cross section alone when the energy (and thus $x$) changes. In Fig.~\ref{ratioplot1} we show ratios of normalized cross sections for 14~TeV and 8.8~TeV or 5~TeV.

\begin{figure}
    \centering
    \includegraphics[width= 0.55\textwidth, height=6.5cm]{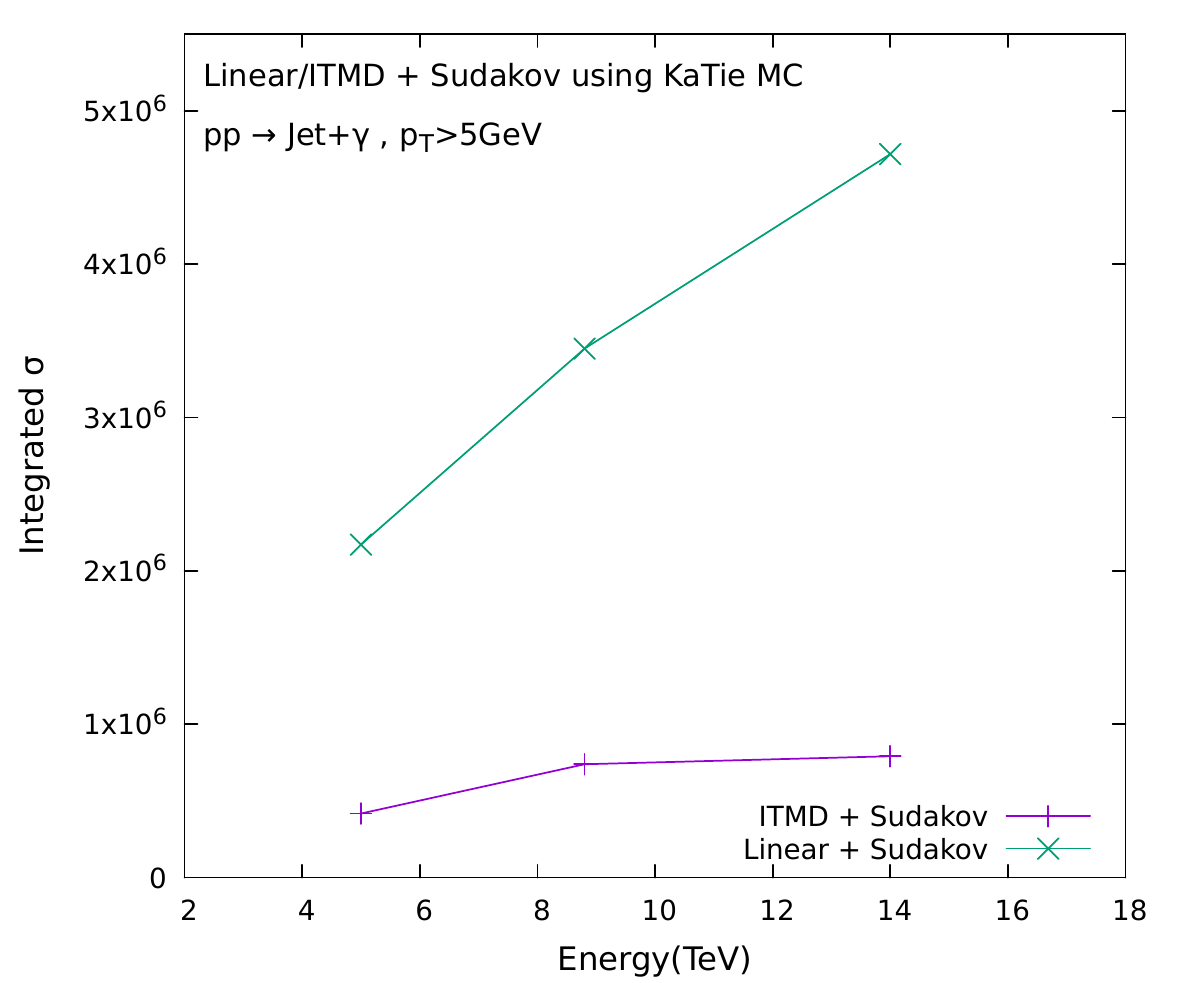}
    \caption{Energy dependence of $p_T$-integrated cross section in pp for ITMD + Sudakov and linear gluon distribution + Sudakov.}
    \label{energydep}
\end{figure}

We observe an interesting pattern. For moderate $p_T$ cuts we observe a growing suppression in the large $\Delta\varphi$ region and enhancement for smaller $\Delta\varphi$. Small transverse momentum cuts signal only slight suppression for $\Delta\varphi\sim 2.3$. 
In order to understand this better we first redo the calculations with KS TMD gluon distribution, but no hard scale evolution due to the Sudakov form factor (second row of Fig.~\ref{ratioplot1}).
We see that now the trend is similar for all $p_T$ cuts, thus we conclude that the Sudakov resummation gives a nontrivial interplay between the shape of the distribution and the cutoff.
Going further, we study the same observable using the linear version of the KS TMD gluon distribution (third row of Fig.~\ref{ratioplot1}). This was obtained in \cite{Kutak:2012rf}. We observe very similar shape of the curves, except they cross the unity at different points for different energy. This in turn reflects the behavior of the cross section with the energy (see Fig.~\ref{energydep} discussed below).
Finally, we study the running coupling BK equation (rcBK) \cite{Albacete:2010sy,Hentschinski:2022rsa } (fourth row of Fig.~\ref{ratioplot1}). Here we see very different shape of the large $p_T$ curve, which is not actually very surprising, as it is known that the raw rcBK equation is valid at rather small $p_T$ values due to missing higher order corrections.

Since there are subtle energy-dependent features in the ratio plots, we can try to visualise them by looking at a more inclusive quantity. In Fig.~\ref{energydep} we show cross sections integrated over $p_T$ from the threshold of 5~GeV using two evolution scenarios: the primary KS nonlinear evolution with Sudakov and linear evolution with Sudakov. We clearly see that the non-linear eqution gives slower growth of the cross section with energy.

The above observables are excellent probes of the complicated dynamics of the TMD PDFs, using only proton-proton cross sections at measurable energies. Indeed, since the cross sections used to produce the ratios are always normalized, the plots at different energy ratios are subtly affected by the interplay of the saturation scale (which is a function of $x$) and the Sudakov hard scale evolution.

\begin{figure}[h]
    \centering
    \begin{subfigure}[h]{0.49\textwidth}
        \includegraphics[width = \textwidth, height = 5.5cm]{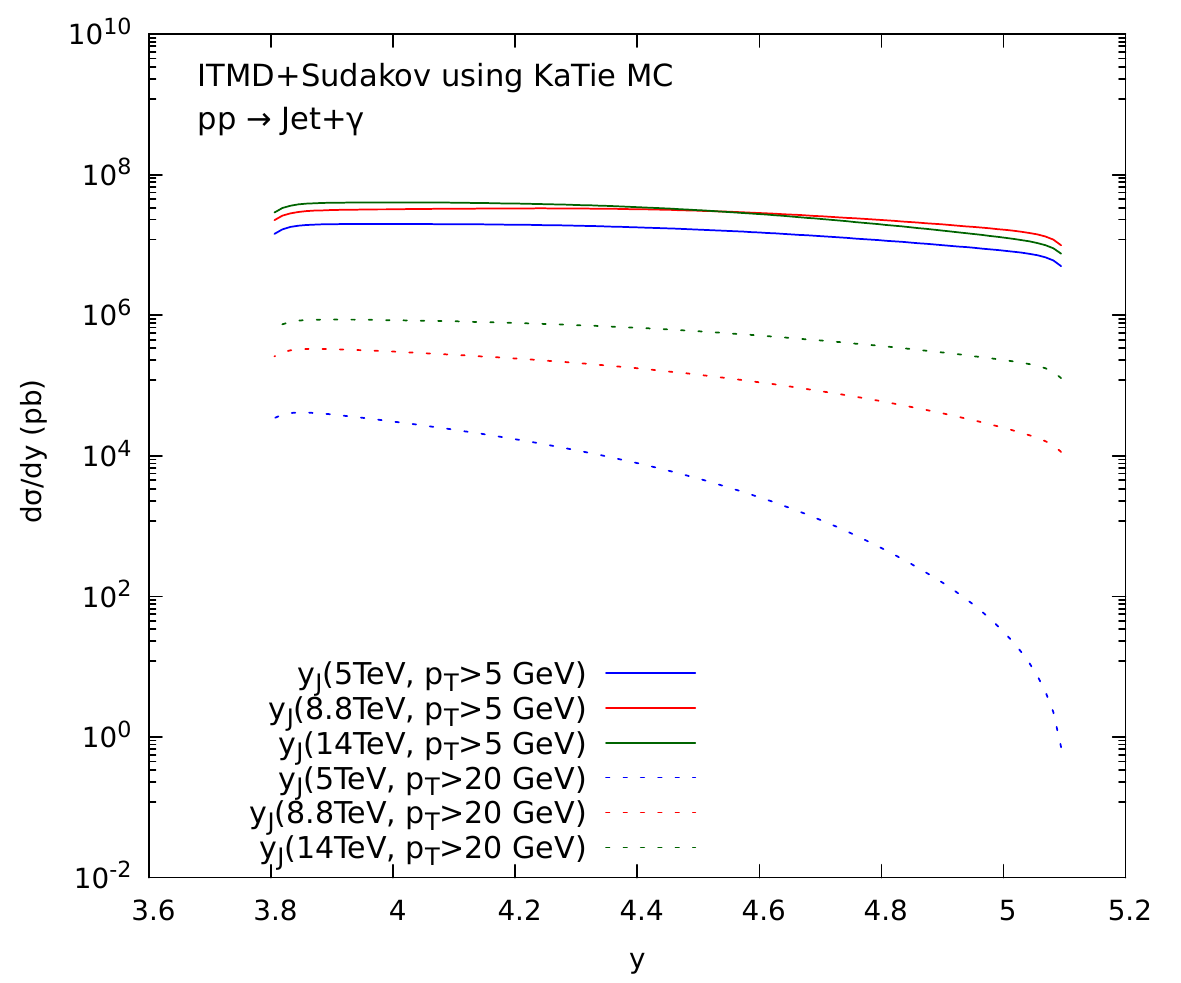}
    \end{subfigure}
    \hfill
    \begin{subfigure}[h]{0.49\textwidth}
        \includegraphics[width = \textwidth, height = 5.5cm]{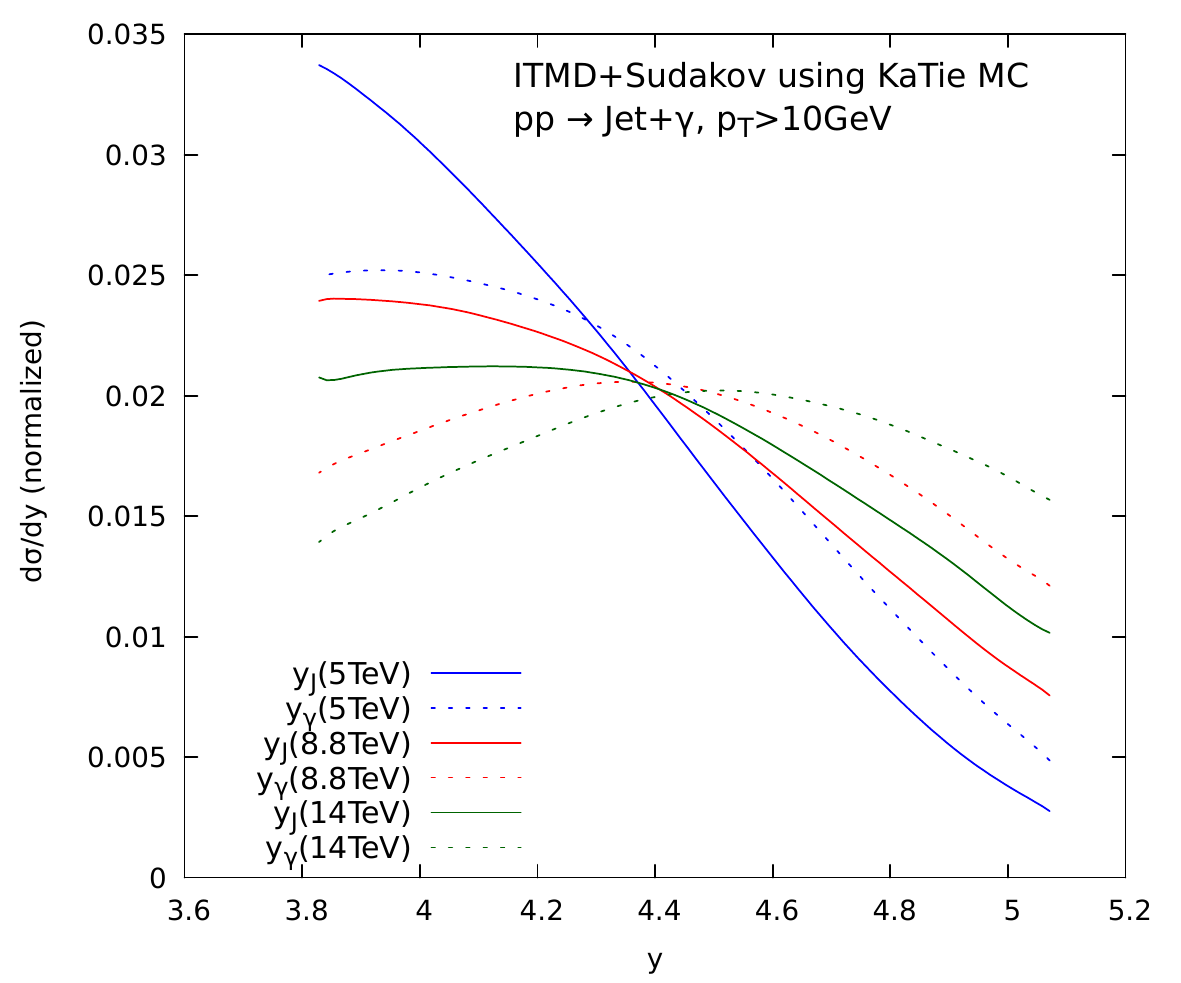}
    \end{subfigure}
\caption{\label{rapidity1} 
Left: Rapidity distributions for different energies and $p_T$ thresholds for jet. Right: Normalized differential cross section for both jet and $\gamma$ at different energies with a fixed $p_T$ threshold of 10~GeV.}
\end{figure}

\begin{figure}
    \centering
    \begin{subfigure}[h]{0.49\textwidth}
    \centering
        \includegraphics[width = \textwidth, height = 5.5cm]{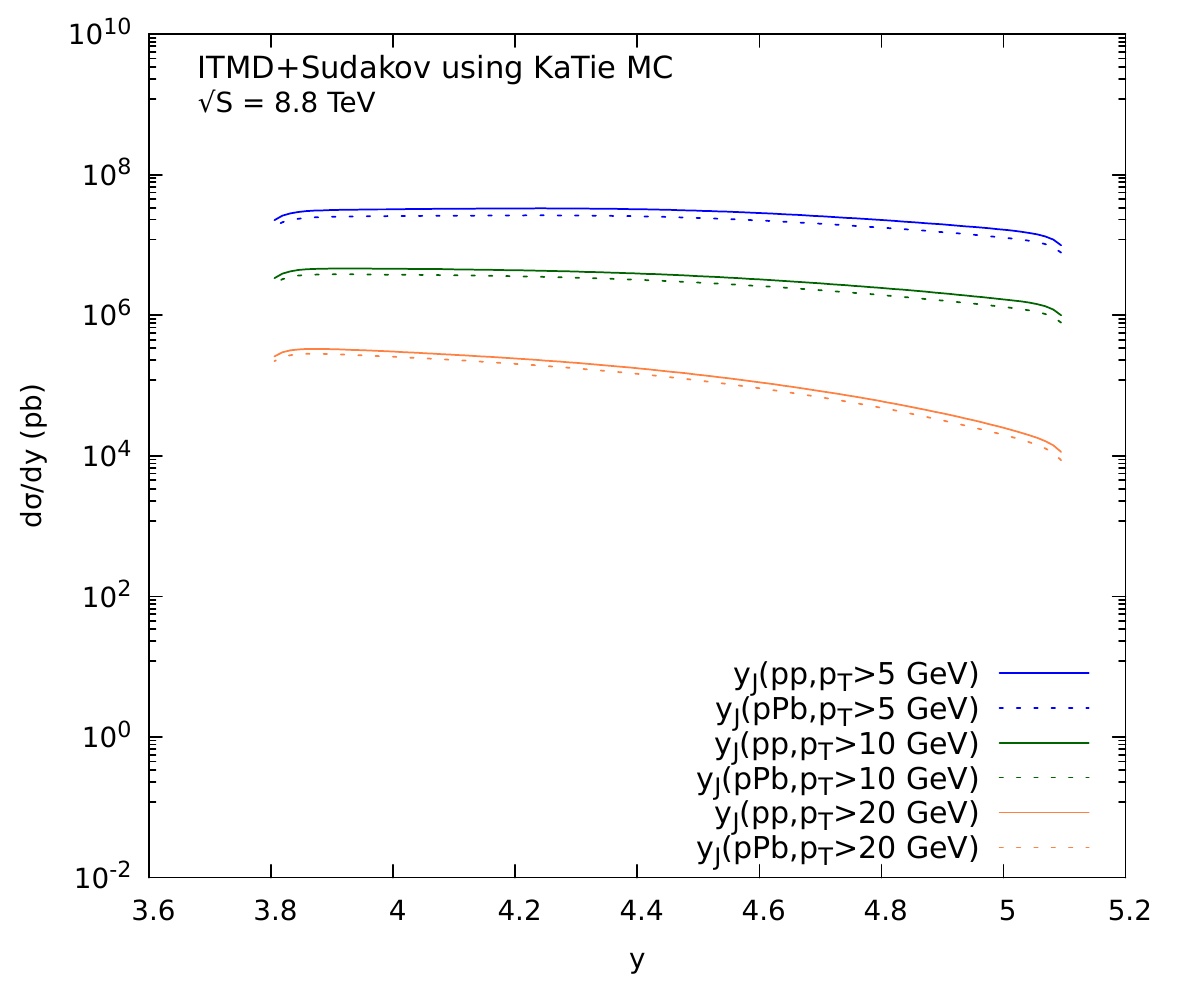}
    \end{subfigure}
    \hfill
    \begin{subfigure}[h]{0.49\textwidth}
    \centering
        \includegraphics[width = \textwidth, height = 5.5cm]{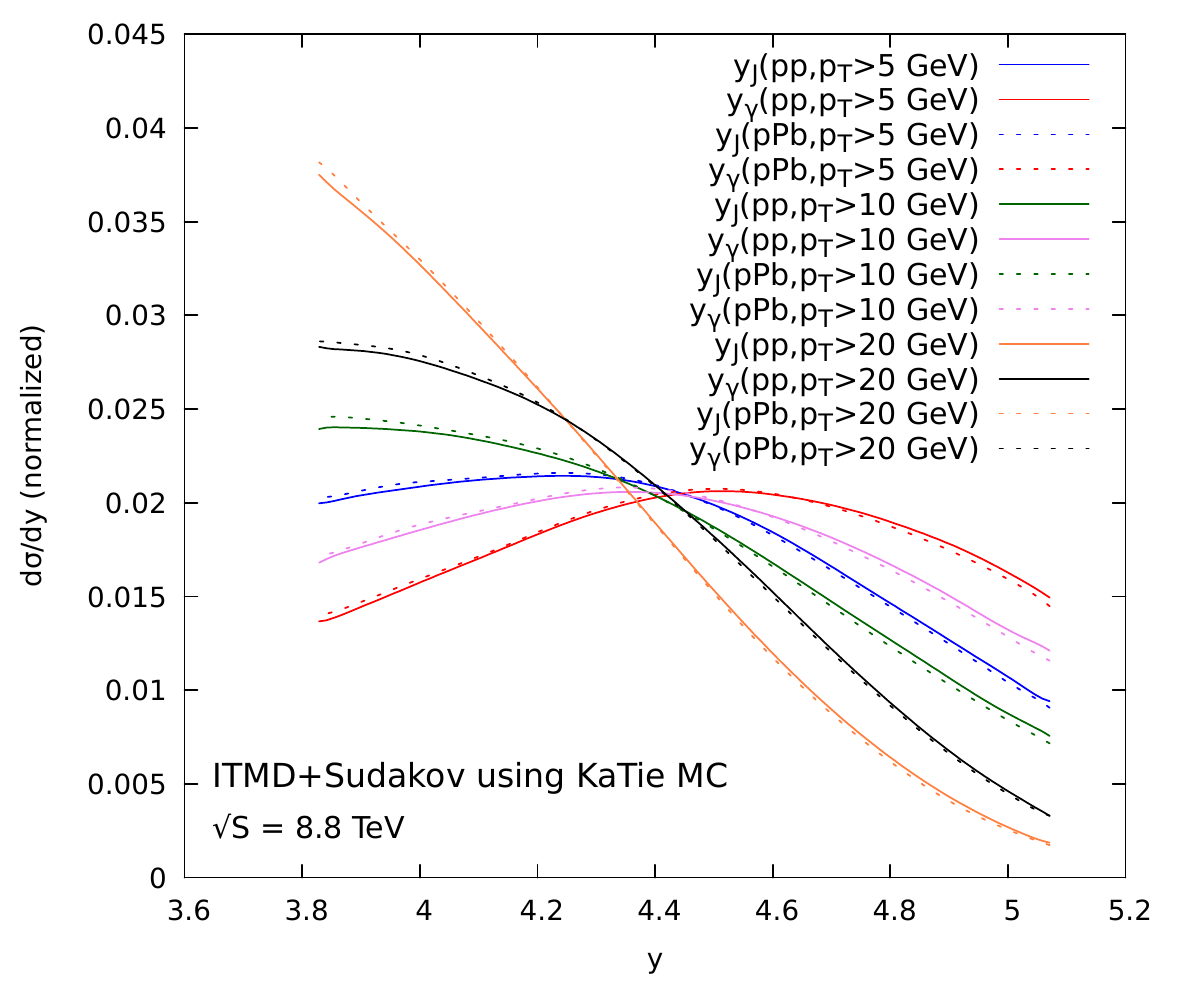}
    \end{subfigure}
\caption{\label{rapidity3}
Rapidity distributions at $\sqrt{s} = 8.8$ TeV for jet (on the left) and for jet and $\gamma$ both at $\sqrt{s} = 8.8$ TeV (on the right) in pp and pPb with different $p_T$ thresholds.}
\end{figure}

\begin{figure}[h]
    \centering
    \includegraphics[width= 0.55\textwidth, height=6.5cm]{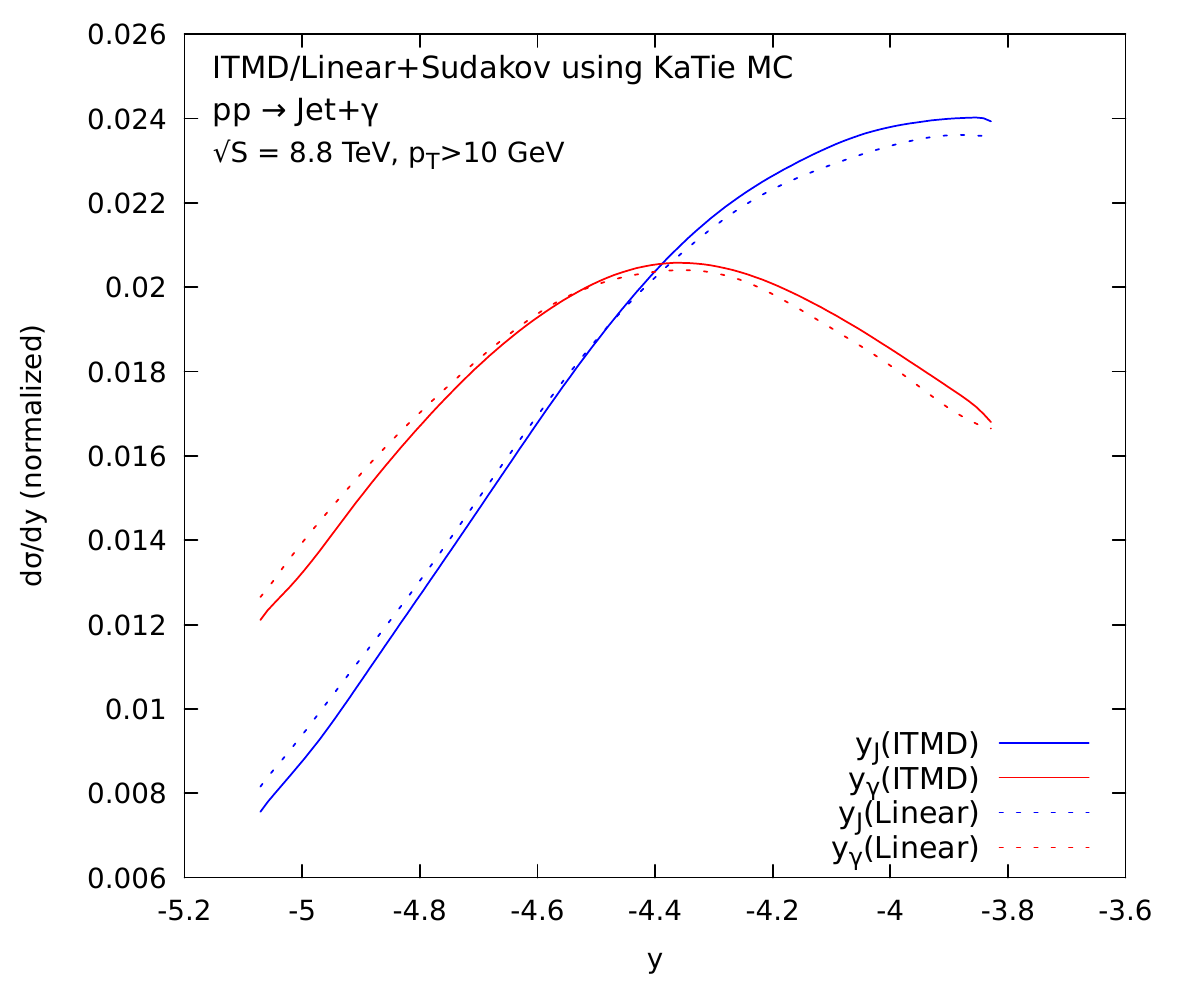}
    \caption{Rapidity distributions at $\sqrt{s} = 8.8$ TeV and a $p_T$ threshold of $10$ GeV for jet and photons in pp with ITMD + Sudakov (bold) and linear gluon distribution + Sudakov (dashed).}
    \label{rapidity_comparison}
\end{figure}

Finally, let us discuss the rapidity distributions.
We start with the p-p collisions at different energies and different $p_T$ cuts (Fig.~\ref{rapidity1}). In the left plot we show differential cross sections as a function of jet rapidity. In the right plot we compare the shapes of the rapidity distributions (normalized plots) for jet and photon at fixed $p_T$ cut of 10~GeV. 
Next, we study dependence of the distributions on the target (Fig.~\ref{rapidity3}). We observe expected suppression due to the saturation mechanism, as well as a slight modification the shape in case of the normalized distributions.
Finally, in Fig.~\ref{rapidity_comparison} we compare normalized rapidity distributions at fixed energy 8.8~TeV and fixed $p_T$ cut of 10~GeV for nonlinear and linear distributions. We observe a slight flattening tendency for the linear gluon distribution, due to the lack of saturation.

\section{Summary}
\label{sec:summary}

The present work was devoted to a detailed study of the forward prompt photon and jet production process at the LHC in the FoCal kinematics. We have used the small-$x$ Improved TMD factorization framework, which is the leading genuine twist approximation to the Color Glass Condensate approach for two particle production processes, within the hybrid (dilute-dense) factorization. An important property of the factorization formula is that it invoves only one TMD gluon distribution -- the dipole TMD gluon distribution, unlike the dijet production in hadron-hadron scattering which inloves five independent distributions at large $N_c$, or dijet production in DIS, which involves the Weizsacker-Williams gluon distribution. The necessary hard scale dependent dipole TMD gluon distribution for proton and lead were obtained in the previous work \cite{Al-Mashad:2022zbq}. Usage of the \KaTie\ Monte Carlo implementing the framework allowed to study various observables: azimuthal correlations, transverse momentum spectra, distributions of the probed longitudinal fractions $x$ and the rapidity distributions. All distributions were computed for proton-proton and proton-lead collisions.  We also studied the ratios of azimuthal correlations just for proton-proton but for different experimentally available energies.

We observe, that the nuclear modification ratios show strong to moderate suppression, depending on the transverse momentum cutoff; for $p_T>5$~GeV we have around 40\% suppression at the back-to-back peak, while for $p_T>20$~GeV we get the suppression of about 20\%.

We also studied evolution of $\gamma$+jet cross section in proton-proton collisions at three available energies: 5.0~TeV, 8.8~TeV and 14~TeV. The convenient way of doing that is to compare shapes of ratios of distributions at different energies. We observe subtle, but very interesting effects within this energy range, that could help to discriminate different high energy evolution scenarios. In similar goal in mind, we studied rapidity distributions.

\section{Acknowledgements}
\label{sec:acknowledgements}
We are grateful I.-C.~Arsene,  P.~Jacobs, C.~Loizides, J.~Otwinowski and T.~Peitzmann for discussions.
IG and PK are supported by the Polish National Science Centre, grant no. DEC-2020/39/O/ST2/03011.
KK acknowledges the Polish National Science Centre grant no. DEC2017/27/B/ST2/01985

\bibliographystyle{JHEP} 
\bibliography{references,references1}

\end{document}